\documentclass[twocolumn,pra,aps,superscriptaddress,showpacs]{revtex4-1}

\usepackage{lineno}
  \usepackage{mathptmx}
\usepackage{subfigure}
\usepackage{dcolumn}
\usepackage{amsmath,amssymb}
\usepackage{bm}
\usepackage{color}
\usepackage{overpic}
\usepackage{latexsym}
\usepackage{epstopdf}
\usepackage{color}
\usepackage[english]{babel}
\usepackage{latexsym}
\usepackage{stmaryrd}

\usepackage{psfrag,graphicx} 
\usepackage{epsf} 
\usepackage{subfigure} 
\usepackage{amsmath} 
\usepackage{amssymb} 
\usepackage{amsfonts}
\usepackage{bm}
\usepackage{natbib}
\usepackage{epstopdf}
\usepackage{appendix}

\definecolor{mygrey}{gray}{0.35}
\definecolor{myblue}{rgb}{0.2,0.2,0.8}
\definecolor{myzard}{cmyk}{0,0,0.05,0}
\definecolor{mywhite}{rgb}{1,1,1}
\definecolor{myred}{rgb}{1,0.,0.3}

\usepackage[colorlinks=true,citecolor=myblue,linkcolor=myred]{hyperref}
\def\be{\begin{equation}}
\def\ee{\end{equation}}
\def\ba{\begin{align}}
\def\enda{\end{align}}
\def\bi{\begin{itemize}}
\def\ei{\end{itemize}}

 \def\ee{\mathord{\rm e}}
 
 \def\ii{\mathord{\rm i}}

\def\half{\textstyle\frac{1}{2}}

\def\fourth{\textstyle\frac{1}{4}}

 \def\ee{\mathord{\rm e}}
 
 \def\ii{\mathord{\rm i}}

\def\half{\textstyle\frac{1}{2}}

\def\fourth{\textstyle\frac{1}{4}}

\renewcommand{\ii}{{\rm i}}
\renewcommand{\ee}{{\rm e}}

\def\beq{\begin{equation}}
\def\beq{\begin{equation}}
\def\eeq{\end{equation}}

 \newcommand{\ket}[1]{|#1\rangle}
 \newcommand{\bra}[1]{\langle #1|}

\begin{document}


\title[Short Title]{Robust Trapped-Ion Quantum Logic Gates   by  Continuous Dynamical Decoupling}

\author{A. Bermudez}
\affiliation{Institut f\"{u}r Theoretische Physik, Albert-Einstein Allee 11, Universit\"{a}t Ulm, 89069 Ulm, Germany}
\author{P. O. Schmidt}
\affiliation{QUEST Institute, Physikalisch-Technische Bundesanstalt and Leibniz University Hannover, 38116 Braunschweig, Germany}
\author{M. B. Plenio}
\affiliation{Institut f\"{u}r Theoretische Physik, Albert-Einstein Allee 11, Universit\"{a}t Ulm, 89069 Ulm, Germany}
\author{A. Retzker}
\affiliation{Institut f\"{u}r Theoretische Physik, Albert-Einstein Allee 11, Universit\"{a}t Ulm, 89069 Ulm, Germany}

\begin{abstract}
We introduce a novel scheme that combines   phonon-mediated quantum logic gates in trapped ions  with the benefits of continuous dynamical decoupling. We demonstrate theoretically that a strong driving of the qubit  decouples it from  external magnetic-field noise, enhancing the fidelity of  two-qubit quantum gates.  Moreover, the scheme does not require ground-state cooling, and is inherently robust to undesired ac-Stark shifts.    The  underlying  mechanism can be extended to a variety of other systems where a strong driving protects the quantum coherence of the qubits without compromising the  two-qubit  couplings.
\end{abstract}

\maketitle

A quantum processor  is an isolated quantum device where information can be stored  quantum-mechanically over long periods of time, but also   manipulated and retrieved.  This forbids its perfect isolation, making such a device sensitive  to the noise introduced by  either external sources,  or experimental imperfections. Additionally, the interactions between distant quantum bits (qubits),  as required to perform quantum logic operations,  are frequently  achieved  by  auxiliary (quasi)particles whose fluctuations introduce an additional source of noise. As  emphasized recently~\cite{qc}, one of the big challenges of quantum-information science is the  quest for   methods to cope with all these natural error sources, achieving error rates that allow fault tolerant quantum error correction.

We address this problem in detail for a prominent   architecture, namely, trapped atomic ions~\cite{WinelandBlatt}. 
Among the most relevant  sources of noise in this system~\cite{wineland_review}, we can list the following: {\it (i)}  thermal noise introduced by  auxiliary quasiparticles (i.e. phonons), {\it (ii)}fluctuating external magnetic fields, {\it (iii)} uncompensated ac-Stark shifts due to fluctuations in the laser parameters, and {\it (iv)} drifts in the phases of the applied laser beams.  There are two different strategies to overcome these obstacles:   {\it (a)} minimize the thermal fluctuations by laser cooling~\cite{cirac_zoller_gate}, searching   for gates operating faster than the timescale set by the other noise sources~\cite{fast_gates},  or {\it (b)} look for schemes that are   {\it intrinsically  robust} to the different types of noise. Among the latter, there are certain schemes that provide partial solutions to the above noise fluctuations, such as  gates that are robust to the thermal motion of the ions~\cite{ms_gate,phase_gate}, or the encoding in magnetic-field insensitive states~\cite{clock_states_long_coherence_times} and decoherence-free subspaces~\cite{decoherence_free}. Recently, there has been a growing effort to implement microwave-based quantum-information processing~\cite{microwave_gates_wunderlich,mw_decoupling,microwave_gates_wineland}, exploiting the excellent control over the phase and amplitude of microwaves as compared to laser fields. Despite  these efforts, it remains a key challenge to suppress all of the  above  sources of noise. Here, we propose to accomplish such a step, achieving fault-tolerant error bounds, by a continuous version of dynamical decoupling at reach of current technology.

While pulsed dynamical decoupling is a well-developed technique~\cite{dynamical_decoupling}  that has already been demonstrated for  ions~\cite{dd_trapped_ions}, its optimal combination  with two-qubit gates requires a considerable additional effort~\cite{decoupling_gates}. Hence, simpler protocols are a subject of recent interest~\cite{Rabl,mw_decoupling}. We hereby  present  a decoupling scheme well suited, but not limited, to trapped-ion experiments with three important properties: {\it  generality,  simplicity,} and {\it  robustness}.  This scheme is sufficiently general  to be applied to any type ion qubits. It is simple since it combines two standard tools  available in most ion trap laboratories, namely a carrier and a red-sideband excitation. In particular, it relies on the strong driving of the carrier transition, which may be realized by laser beams for optical qubits, or by  microwaves  for hyperfine and Zeeman qubits.  With this independent driving source, we  improve simultaneously  the performance and the speed of the gate,  as compared to the light-shift gates~\cite{light_shift_gates}. Besides,  this driving has the potential of simplifying certain aspects of previous gate schemes~\cite{supp_mat}, and is responsible for the  gate robustness   at different levels. On top of decoupling the qubits from the magnetic-field noise, it suppresses the errors due to the thermal ion motion, and to uncompensated ac-Stark shifts. When focusing on hyperfine or Zeeman qubits, we can further benefit from the phase and amplitude stability of microwaves.  Alternatively, concatenated drivings may overcome amplitude fluctuations~\cite{cai}.     

{\it The system.--} We focus on $^{25}$Mg$^{+}$ to exploit the benefits of microwave technology~\cite{magnesium}, although the  scheme is also valid for  other  ion species. Let us consider two hyperfine states $\ket{0},\ket{1}$ with energy difference $\omega_0$ to form our qubit (see Fig.~\ref{dipole_forces}{\bf (a)} and Table~\ref{o_magnitude}). The ions arrange in a  string along the axis of  a linear Paul trap characterized by  the radial and axial frequencies $\omega_{ x}, \omega_{ z}$ (Table~\ref{o_magnitude}). The  small radial vibrations  yield a set of collective vibrational modes of frequencies $\omega_n$, whose excitations are the transverse  phonons $a_n,a_n^{\dagger}$~\cite{supp_mat}. If the qubit-qubit couplings  are mediated by these  quasiparticles~\cite{transverse_phonon_gates,ising_porras}, the scheme becomes less sensitive to ion-heating/thermal motion, and  easier to operate within the Lamb-Dicke regime. 

 As shown in Fig.~\ref{dipole_forces}{\bf (a)}, a pair of laser beams in a Raman configuration (red arrows)  induces a transition between the qubit states via an auxiliary excited state. By setting their frequency beatnote $\omega_{\rm L}$ close to $\omega_0-\omega_n$, such that the detuning $\delta_n=\omega_{\rm L}-(\omega_0-\omega_n)$ is much smaller than the radial trap  frequency (see Table~\ref{o_magnitude} for the bare detuning $\delta_{\rm L}=\omega_{\rm L}-(\omega_0-\omega_x)$), one obtains the  red-sideband excitation.  In addition, we  drive the carrier transition (blue arrow). For our particular qubit choice, this driving can be performed with  microwave radiation  of frequency $\omega_{\rm d}$, such that the complete Hamiltonian is
\begin{equation}
\label{sideband}
H_{\rm c}+H_{\rm r}=\sum_i\frac{\Omega_{\rm d}}{2}\sigma_i^+\ee^{-\ii(\omega_{\rm d}-\omega_0)t}+\sum_{in}\mathcal{F}_{in}\sigma_i^{+}a_n^{\phantom{\dagger}}\ee^{-\ii \delta_nt}+\text{H.c.},
\end{equation}
where we have introduced the microwave Rabi frequency $\Omega_{\rm d}$, and the sideband coupling strengths $|\mathcal{F}_{in}|\propto\Omega_{\rm L}\eta$ scale  linearly with the laser Rabi frequency $\Omega_{\rm L}$ and Lamb-Dicke parameter $\eta$,  such that $|\mathcal{F}_{in}|\ll\delta_n$ (Table~\ref{o_magnitude}). Here, we use the spin operators  $\sigma_{i}^{+}=\ket{1_i}\bra{0_i}$, and we work in the interaction picture rotating with the phonon and qubit frequencies.

\begin{figure}
\centering
\includegraphics[width=\columnwidth]{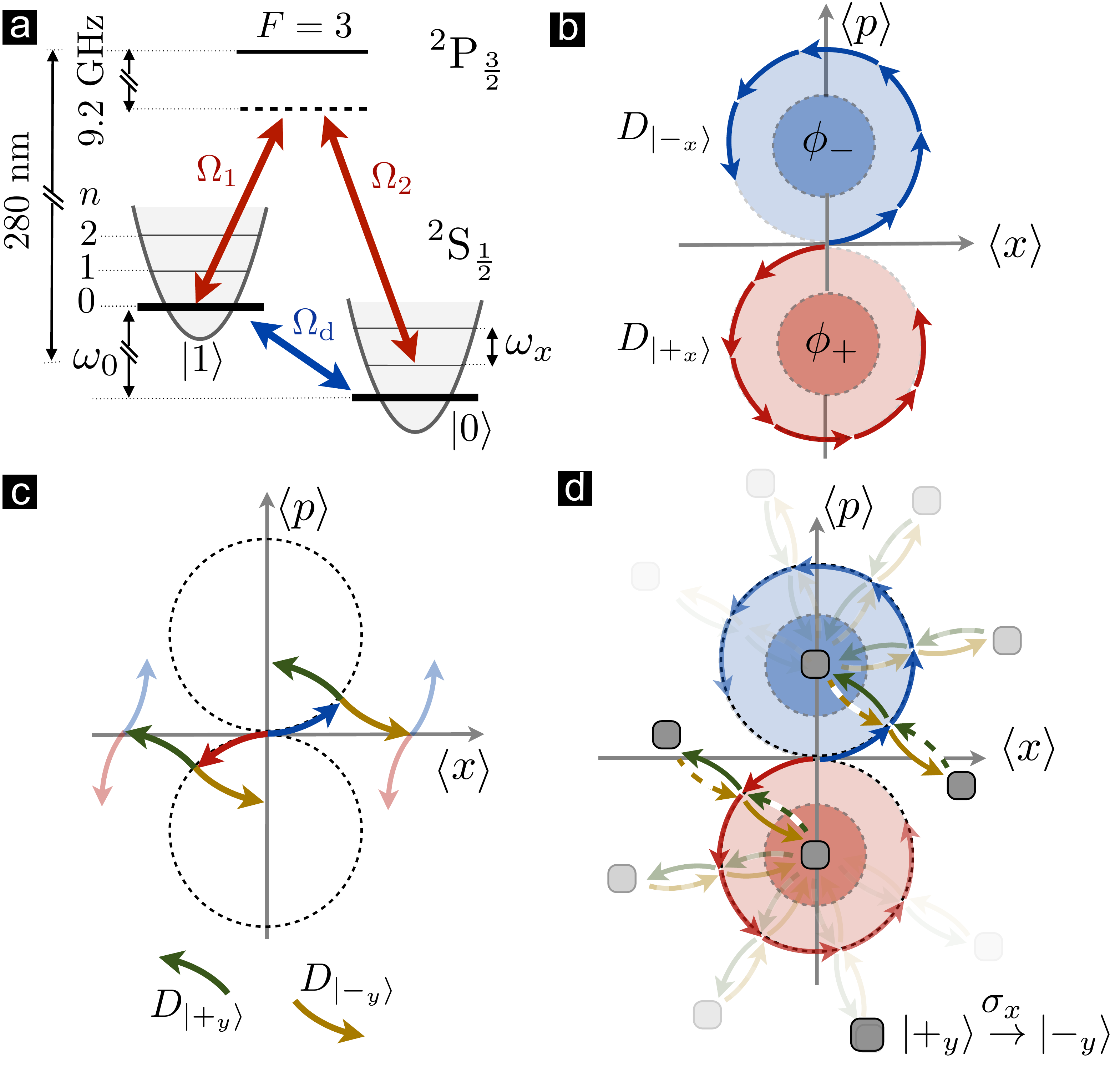}
\caption{ {\bf State-dependent dipole forces:}  {\bf (a)} Hyperfine structure of $^{25}{\rm Mg}^+$. The states $\ket{0}=\ket{F=3,m_{F}=3}$ and $\ket{1}=\ket{2,2}$ from the groundstate manifold $^2{\rm S}_{1/2}$ form our qubit. Two laser beams $\Omega_1,\Omega_2$ drive the red sideband via an off-resonant excited state, and a microwave $\Omega_{\rm d}$ directly couples to the transition.   {\bf (b)} Spin-dependent $\sigma^x$-force acting on a single trapped ion.  The phonons associated to the states $\ket{+_x}$,$\ket{-_x}$ are displaced in phase space according to $D_{\ket{+_x}}(\Delta t)$, $D_{\ket{-_x}}(\Delta t)$, and form different closed paths that lead to the geometric phases $\phi_{\pm}$. {\bf (c)} Trotterization of the combined $\sigma^x$ and $\sigma^y$ forces. The $\sigma^x$-displacement $D_{\ket{+_x}}(\Delta t)$ shall be followed by the two possible $\sigma^y$ displacements $D_{\ket{\pm_y}}(\Delta t)$ since $\ket{+_x}\propto (\ket{+_y}+\ii\ket{-_y}$). Hence, the phase-space trajectory is not generally closed. {\bf (d)} Schematic spin-echo refocusing of the $\sigma^y$-force. By applying a $\pi$-pulse  $X^{\pi}_i=\sigma^x_i$ (grey box) at half the $\sigma^y$-displacements, one obtains $\ket{\pm_y}\to\ket{\mp_y}$, such that the  displacements $D_{\ket{\pm}_y}$ are reversed (dotted arrows), and  the trajectory is refocused yielding a closed path with a well-defined  geometric phase robust to the thermal fluctuations.  }
\label{dipole_forces}
\end{figure}

{\it Sideband gates and thermal noise.--}  We first introduce an intuitive picture to understand the effects of the thermal noise in terms of the geometric phase gates~\cite{ms_gate,phase_gate}. The   spin-boson Hamiltonian~\eqref{sideband} in the absence of the driving $\Omega_{\rm d}=0$ can be expressed  as the combination of two non-commuting spin-dependent forces. Each force aims at displacing the normal modes along a closed trajectory in phase space, which is   determined by the particular eigenstates $\ket{\pm_x},\ket{\pm_y}$ of the  Pauli matrices $\sigma^x,\sigma^y$~\cite{supp_mat}. In Fig.~\ref{dipole_forces}{\bf (b)}, we describe the action of the $\sigma^x$-force 
on a single trapped ion. Depending on the spin state $\ket{\pm_x}$, the ion follows a different path in phase space. After returning to the starting point,  the ion picks a geometric phase  independent of the motional state, and only determined by the area enclosed by the trajectory. The spin dependence of these trajectories, together with the collective nature of the phonons, are the key ingredients of  the two-qubit geometric phase gates for trapped ions in thermal motion. In contrast, for the much simpler single red-sideband, $\sigma_x$ and $\sigma_y$ forces are implemented, resulting in rotations around two orthogonal axis as shown in Fig.~\ref{dipole_forces}{\bf (c)}. In a Trotter decomposition, the consecutive concatenation of these orthogonal infinitesimal displacements spoils the closed character of the trajectory. Since the vibrations do not return to the initial state, the qubit  and phonon states are generally entangled, and the gate becomes sensitive to thermal fluctuations. This intuitive picture of the thermal noise is corroborated below by means of analytic and numerical arguments.

\begin{table}
  \centering 
   \caption{{\bf Specific values of trapped-ion setup}  }
  \begin{tabular}{ c  c c c c c c c}
\hline
\hline
$\omega_0/2\pi$ & $\omega_x/2\pi$ & $\omega_z/2\pi$ & $\eta$ & $|\delta_{\rm L}|/2\pi$ & $\Omega_{\rm L}/2\pi$ & $\Omega_{\rm d}/2\pi$& $B_0$\\
\hline
$1.8 \text{ GHz}$ & $4\text{ MHz}$ & $1 \text{ MHz}$ & 0.2 & 800 kHz &  500 kHz & 5.2 MHz& 4 mT\\
\hline
\hline
\end{tabular}
  \label{o_magnitude}
\end{table}

  For large detunings $|\delta_{\rm L}|\gg\Omega_{\rm L}\eta$, the lasers only excite virtually the vibrational modes, and the phonons can be  adiabatically eliminated. In fact, it is the process where a phonon is virtually created by an ion, and then reabsorbed by a distant one, which gives rise to the effective  couplings
\begin{equation}
\label{xy_model}
H_{\rm eff}=\sum_{ij} J^{\rm eff}_{ij}\sigma_i^+\sigma_j^-,\hspace{1.5ex} J^{\rm eff}_{ij}=-\sum_n\frac{1}{\delta_n}\mathcal{F}_{in}\mathcal{F}_{jn}^*,
\end{equation}
which can be exploited to perform the phonon-mediated  gates, or to implement a quantum simulation  of the  XY spin chain~\cite{xy_model}. At certain instants of time, the  unitary evolution corresponds to a SWAP gate~\cite{nielsen_chuang}, which performs the logic operation $\ket{1_i0_j}\leftrightarrow\ket{0_i1_j}$ while leaving  the remaining states unchanged.  However, there is an additional process ignored in the above explanation that  spoils the performance of the gate, namely, the phonon might be reabsorbed by the same ion. This leads to a    residual spin-phonon coupling  
\begin{equation}
\label{dephasing}
H_{\rm res}=\sum_i\hat{B}_i(t)\sigma_i^{z},\hspace{1ex}\hat{B}_i(t)=\sum_{nm}B_{inm}a_m^{\dagger}a_n^{\phantom{\dagger}}\ee^{-\ii(\omega_n-\omega_m)t},
\end{equation}
where $B_{inm}=-\half\mathcal{F}_{in}\mathcal{F}_{im}^*(\frac{1}{\delta_n}+\frac{1}{\delta_m})$. According to this expression, the ion resonance frequency  fluctuates in time due to the collective motional dynamics. This effect can be interpreted as a {\it local thermal noise} in the limit of many ions, where it leads to dephasing. In Fig.~\ref{swap_gate}{\bf (a)},  the critical effect of this term on the SWAP gate has been studied. We show the results of a numerical simulation of  the time evolution under the full  spin-phonon Hamiltonian, and compare it to the effective idealized description~\eqref{xy_model}.  As  evidenced in this figure, the performance of the gate is severely modified by the thermal phonon ensemble. In fact, the swapping probabilities  only approach the effective description (circles and squares) for ground-state cooled ions.   As the mean phonon number is increased, the oscillations get a stronger damping, and the generation of Bell states at half the SWAP periods (black arrows) deteriorates.

\begin{figure}
\centering
\includegraphics[width=\columnwidth]{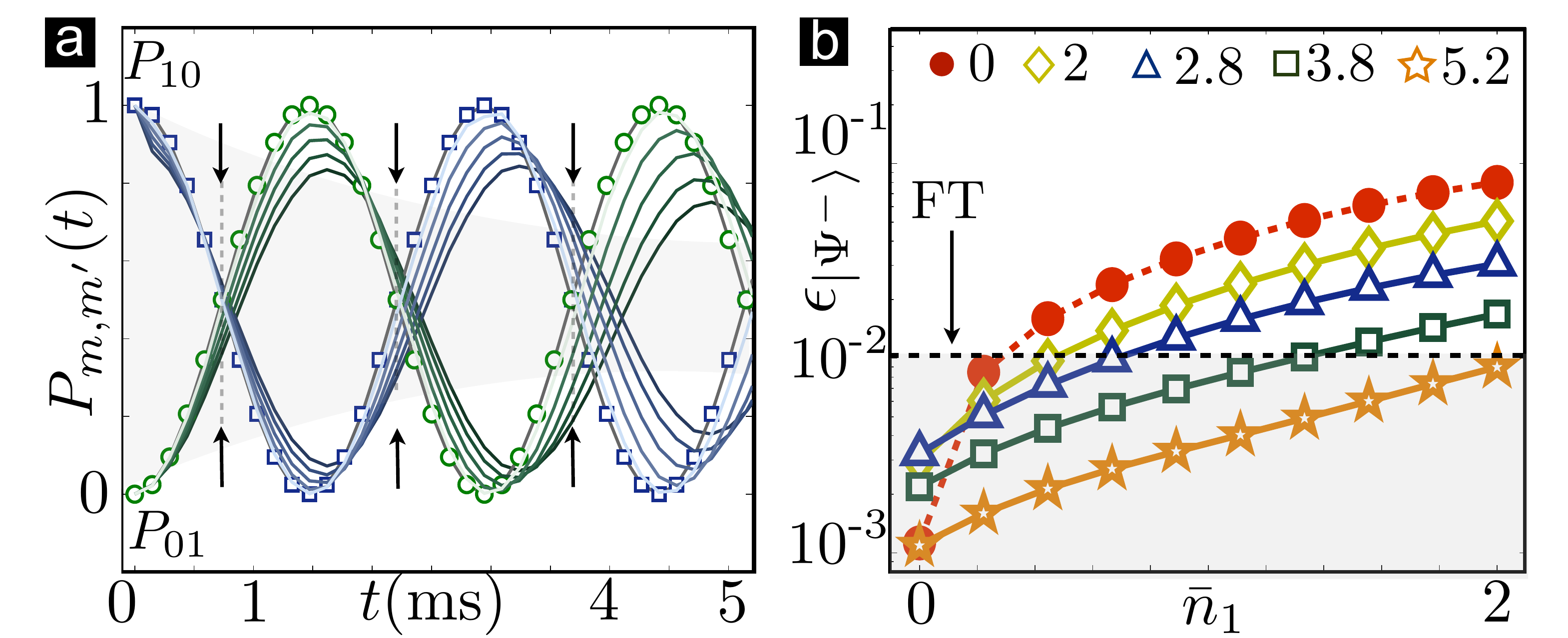}
\caption{ {\bf Robustness with respect to thermal noise:}  {\bf (a)} Dynamics of the swap probabilities $P_{10}(t)$ (squares), $P_{01}(t)$ (circles) for the effective gate~\eqref{xy_model}, compared to the exact spin-phonon Hamiltonian  for a two-ion crystal with different mean phonon numbers $\bar{n}=\{0, 0.1,1,2,4\}$ (solid lines). The phonon Hilbert  space is truncated to $n_{\rm max}=20$ excitations per mode. {\bf (b)} Error $\epsilon=1-\mathcal{F}$ for the generation of the Bell state $\ket{\Psi^{-}}$ from  $\ket{\psi_0}=\ket{10}$ by the driven entangling gate as a function of the mean phonon number and setting $n_{\rm max}=14$. As the driving power is increased, $\Omega_{\rm d}/\omega_z\in\{0,2,3.8,5.2\}$, the fidelity approaches unity. We represent the gate error and compare it to the fault-tolerance (FT) threshold $\epsilon_{\rm t}\sim 10^{-4}-10^{-2}$. }
\label{swap_gate}
\end{figure}

{\it Achieving robustness against thermal noise.--} We now address the question of protecting the coherent spin dynamics from this thermal dephasing by switching $\Omega_{\rm d}\neq 0$. Schematically, this may be accomplished by refocusing the effects of one of the spin-dependent forces using a series of spin-echo pulses that invert the atomic state~[Fig.~\ref{dipole_forces}{\bf (d)}]. Note that the dynamics of the dephasing~\eqref{dephasing} is characterized by the difference between  normal-mode frequencies,  $|\omega_n-\omega_m|\leq 2(\omega_z^2/\omega_x)\ll \omega_z$, and  can thus be  considered as a low-frequency noise.  Unfortunately, this noise is still much faster than the typical gate times, and simple spin-echo techniques ~\cite{spin_echo} would not suffice to get rid of the thermal  errors. Instead of using complicated pulse sequences, we show below that the strong   driving  of the carrier transition $\ket{0}\leftrightarrow\ket{1}$ implements a continuous version of the refocusing of Fig.~\ref{dipole_forces}{\bf (d)}, providing a viable and simple mechanism for overcoming this noise.  
As  will become clear later on,  this  driving is not only responsible for the decoupling  from the thermal noise, but it  also minimizes the  undesired errors due to ac-Stark shifts and magnetic-field noise. 

A helpful account  of the  decoupling mechanism may be the following. In the dressed-state basis of the driving   $\ket{\pm_x}_i=(\ket{1_i}\pm\ket{0_i})/\sqrt{2}$,  the residual  spin-phonon coupling becomes
\begin{equation}
\label{dephasing_dressed}
H_{\rm res}(t)=\sum_{inm}B_{inm}\ket{+_x}_i\bra{-_x}a_m^{\dagger}a_n^{\phantom{\dagger}}\ee^{\ii(\Omega_{\rm d}-(\omega_n-\omega_m))t}+\text{H.c.}
\end{equation}
For a strong driving strength $\Omega_{\rm d}$, this term  rotates very fast even for two vibrational modes that are close in frequency, and  can be thus neglected in a rotating wave approximation. Note that we have assumed a vanishing phase of the driving, but the same argument applies for any other stable phase~\cite{supp_mat}.

  This simple argument has to be readdressed  for a combination of the carrier and red-sideband interactions~\eqref{sideband}, since the residual couplings are no longer described by Eq.~\eqref{dephasing}. In order to show that a similar argument can still be applied,  we have performed a polaron-type transformation that allows us to find the effective interaction and residual error terms to any desired order of perturbation theory~\cite{supp_mat}. We find that the dynamics is accurately described by the effective Hamiltonian
\begin{equation}
\label{xx_model}
\tilde{H}_{\rm eff}=\sum_{ij} \tilde{J}^{\rm eff}_{ij}\sigma^x_i\sigma^x_j+\half\sum_i\Omega_{\rm d}\sigma^x_i,\hspace{1ex}\tilde{J}^{\rm eff}_{ij}=\fourth J^{\rm eff}_{ij},
\end{equation}
whereas the residual spin-phonon coupling is given by
\begin{equation}
\label{res}
\tilde{H}_{\rm res}=\sum_{in}\frac{\ii}{2}\left(\mathcal{F}_{in}a_n^{\phantom{\dagger}}-\mathcal{F}_{in}^*a_n^{\dagger}\right)\left(\text{cosh}\hat{\Theta}_i\sigma_i^y-i\text{sinh}\hat{\Theta}_i\sigma_i^z\right)
\end{equation}
such that $\hat{\Theta}_i=\sum_m(\mathcal{F}_{im}a_m^{\phantom{\dagger}}-\mathcal{F}_{im}^*a_m^{{\dagger}})/2\delta_m.$ By moving to the dressed-state basis, the residual  term  only involves transitions between the dressed eigenstates $\ket{+_x}\leftrightarrow\ket{-_x}$, supplemented by the transformation on the phonons encoded in the different powers of $\hat{\Theta}_i$. Fortunately, all these transitions are inhibited due to the large energy gap between the dressed states   set by $\Omega_{\rm d}$. More precisely, in the strong driving  regime $\Omega_{\rm d}\gg2\delta_n$ (see Table~\ref{o_magnitude}),  the leading order terms of the residual coupling~\eqref{res} can be neglected in a rotating wave approximation. 

To check the correctness of this argument, we integrate numerically the complete Hamiltonian with  both the sideband and the carrier terms~\eqref{sideband}, and take into account the thermal motion of the trapped ions.   After the unitary evolution $U(t_{\rm f})=U(t_{\rm f},\half t_{\rm f})(\sigma_i^z\sigma_j^z)U(\half t_{\rm f},0)$, we calculate the fidelity of  producing the Bell state $\ket{\Psi^{-}}=(\ket{10}-\ii\ket{01})/\sqrt{2}$, $\mathcal{F}_{\ket{\Psi-}}={\rm max}_{t_{\rm f}}\big\{\langle \Psi^-|\text{Tr}_{\text{ph}}\{U^{\dagger}(t_{\rm f})\rho_0U(t_{\rm f})\}|\Psi^-\rangle\big\}$ for different initial thermal states and  driving strengths.  The results displayed in Fig.~\ref{swap_gate}{\bf (b)} demonstrate the promised decoupling from the thermal noise.  We observe that the fidelity of the gate improves considerably with respect to the non-driven case when the amplitude lies beyond    $\Omega_{\rm d}\gg 2\delta_n$.  For driving strengths in the 4-5 MHz range, the gate error  lies within the fault-tolerance threshold $\epsilon_{\rm t}\sim 10^{-2}-10^{-4}$~\cite{threshold} for  states with  mean phonon numbers  $\bar{n}\leq 2$ (see inset). However, we emphasize that a lower fidelity should be expected when experimental imperfections are taken into account.

In Fig.~\ref{decoupled_gate_scheme}, we describe  the experimental steps required  to create the desired Bell state.
In the first step, the qubits are optically pumped to $\ket{0}$, and the initial state $\ket{\psi_0}=\ket{10}$ is prepared by a $\pi$-pulse $X_1^{\pi}$ obtained from a microwave resonant with the carrier transition (blue). Three comments are now in order. First, a  magnetic field $B_0\approx 4$ mT needs to be applied to ensure that the Zeeman splitting between magnetic states is sufficiently large to avoid driving unwanted transitions. Note that the motional excitation of the ions can be neglected when driving microwave transitions owing to the vanishing Lamb-Dicke factor. Second, either the ac-Stark shift from an off-resonant  laser beam or a magnetic field gradient  is required to effectively hide the second ion. Alternatively, one could use ion shuttling techniques~\cite{shuttling}. Third, we account for the worst possible scenario by considering {\it (i)} different switching times of the lasers and  microwaves, and  {\it (ii)} imperfect timing with the microwave. By introducing  global $\pi$-pulses $Z_1^{\pi}Z_2^{\pi}$ from the energy shift of an off-resonant strong microwave (yellow), we refocus the fast oscillations caused by the resonant microwave and correct the possible difference of switching times. In the second step, the two-qubit coupling is applied  at  $t=t_0$ by switching on the laser beams responsible for the red sideband. Again, a refocusing  pulse  at $t=t_{\rm f}/2$ shall correct for the imperfect synchronization  with the resonant microwave. Let us stress that these  pulses may not be required in the case of accurate synchronization. In the final step, after switching off the laser and microwaves at $t=t_{\rm f}$,  the qubit state is measured by state-dependent fluorescence techniques. If the announced decoupling has worked correctly, this two-qubit gate should have generated the entangled Bell state
$\ket{\Psi^-}$ regardless of the phonon state. Let us  emphasize that  this gate is capable of producing  the remaining Bell states by choosing different initial states~\cite{supp_mat}   and, together with single-qubit rotations, becomes universal for quantum computation.

\begin{figure}
\centering
\includegraphics[width=\columnwidth]{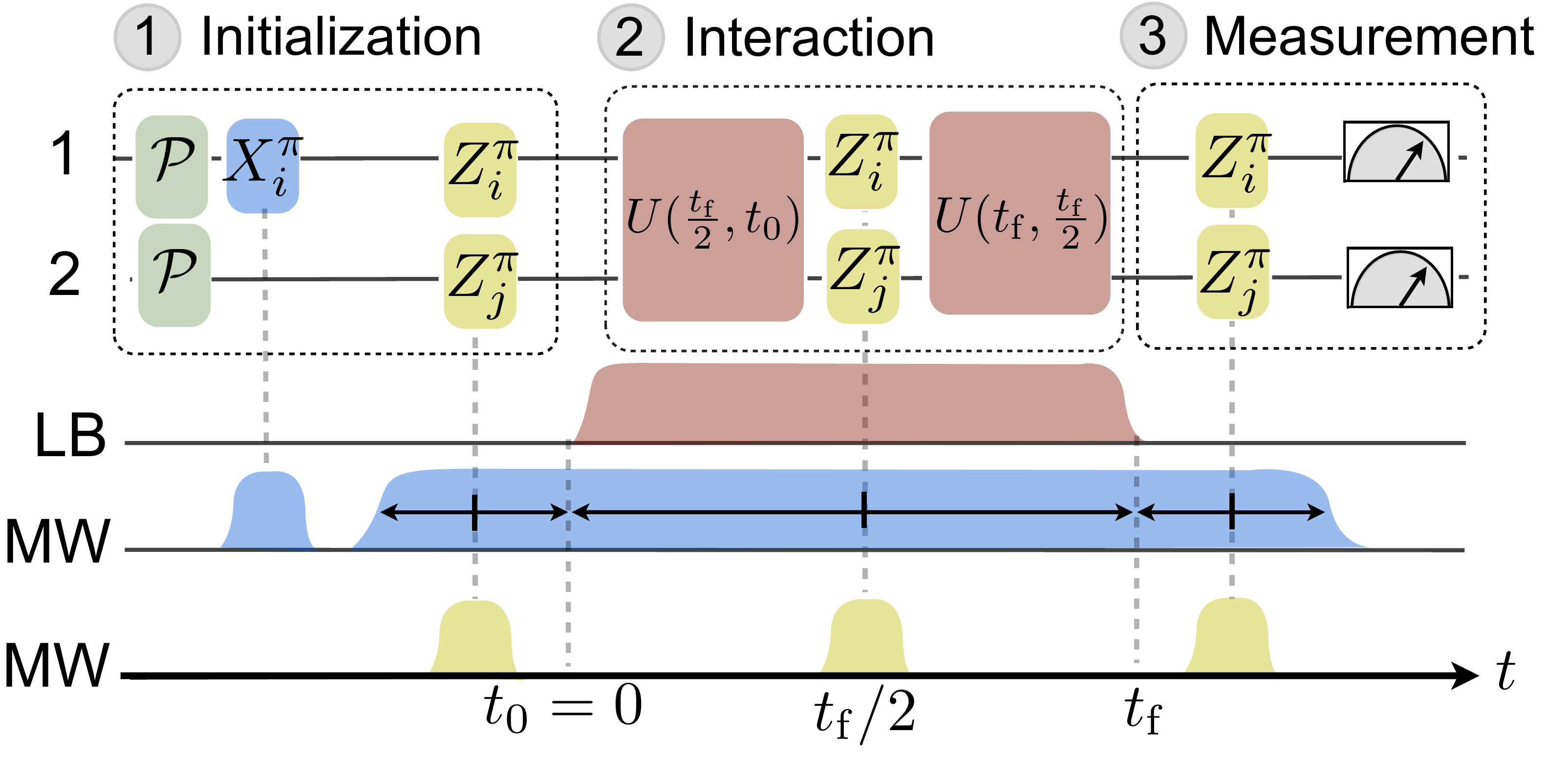}
\caption{ {\bf Scheme for the creation of entangled states:} The two upper rows represent the circuit model for the two qubits, and the lower rows represent the switching of the laser beams (LB) and microwaves (MW)  responsible of driving the sideband and carrier transitions.  The initialization consists of optical pumping $\mathcal{P}$, followed by a local $\pi$-pulse $X_i^\pi=\sigma_i^x$. The global $\pi$-pulses $Z^{\pi}_i=\sigma_i^z$ correct the possible synchronization errors. The two-qubit coupling  is applied by switching on the laser beams responsible for the red-sideband (red), and the microwaves that yield the noise decoupling (blue).  }
\label{decoupled_gate_scheme}
\end{figure}

{\it Resilience to magnetic-field noise and ac-Stark shifts.--} So far, we have neglected  the effects of the environment. In standard traps, the leading source of noise is due to environmental fluctuating magnetic fields, which limit the coherence times of magnetic-field sensitive states to milliseconds~\cite{wineland_review}. This is particularly important for multi-ion entangled states experiencing super-decoherence~\cite{gates_decoherence}, and will also play a key role in our scheme considering that the  two-qubit couplings lie in the $\tilde{J}_{\rm eff}/2\pi\approx 1$ kHz regime. To study its consequences, we  model  the {\it global magnetic-field noise} by a fluctuating resonance frequency $H_{\rm n}=\half\sum_iF(t)\sigma_i^z$. Here, $F(t)$ is a stochastic Markov process~\cite{random_processes,ou_process} characterized by a diffusion constant $c$  and a correlation time $\tau$ that determine the exponential decay of the   coherences $\langle\sigma^x(t)\rangle=\ee^{-t/T_2}$ with $T_2=2/c\tau^2$~\cite{supp_mat}. By fixing these parameters, we   can reproduce the experimentally observed $T_2\approx5$ms [Fig.~\ref{decoupled_gate}{\bf (a)}], and study its consequences on the two-qubit entangling gate. Notice that  the evolution within the zero-magnetization subspace is not affected by this global noise, which can be interpreted as a decoherence-free subspace.  Hence, we have studied the fidelity  of generating a Bell state that lies  outside this subspace $\ket{\Phi^-}=(\ket{11}-\ii\ket{00})/\sqrt{2}$. 

As shown below, the strong  driving $\Omega_{\rm d}$ protects the qubit coherences from this magnetic noise without compromising  the entangling gate, and may be considered as a continuous version~\cite{continuous_decoupling,mw_decoupling} of the so-called dynamical decoupling~\cite{dynamical_decoupling}.   To single out the effects of the  noise from those of the thermal motion, we have considered a ground-state cooled crystal,  setting $\Omega_{\rm d}/2\pi=5.2$MHz to ensure that the results can be carried out to higher temperatures [Fig.~\ref{swap_gate}{\bf (b)}]. We evaluate numerically the fidelity of generating  the Bell state $\ket{\Phi^-}$  by  averaging over different samplings of the random noise [inset of Fig.~\ref{decoupled_gate}{\bf (b)}]. Due to the decoupling, the fidelity approaches unity at the gate time $t_{\rm f}=0.7$ ms. In the main panel of Fig.~\ref{decoupled_gate}{\bf (b)}, we show that the gate error   for shorter coherence times still lies below the fault-tolerance threshold, which demonstrates that the decoupling mechanism supports a stronger magnetic noise. Alternatively, this  tells us that the gate  tolerates smaller speeds, and thus  lower  Rabi frequencies of the Raman beams. This shall reduce even further the thermal error studied above, and the spontaneous scattering of photons due to the Raman beam configuration in Fig.~\ref{dipole_forces}{\bf (a)}. We have also calculated the fidelity of the quantum channel with respect to the desired quantum gate $U_{\rm eff}=\ee^{-\ii\frac{\pi}{4}\sigma_i^x\sigma_j^x}$~\cite{supp_mat}, which also lies  within the fault-tolerance threshold.

\begin{figure}
\centering
\includegraphics[width=1\columnwidth]{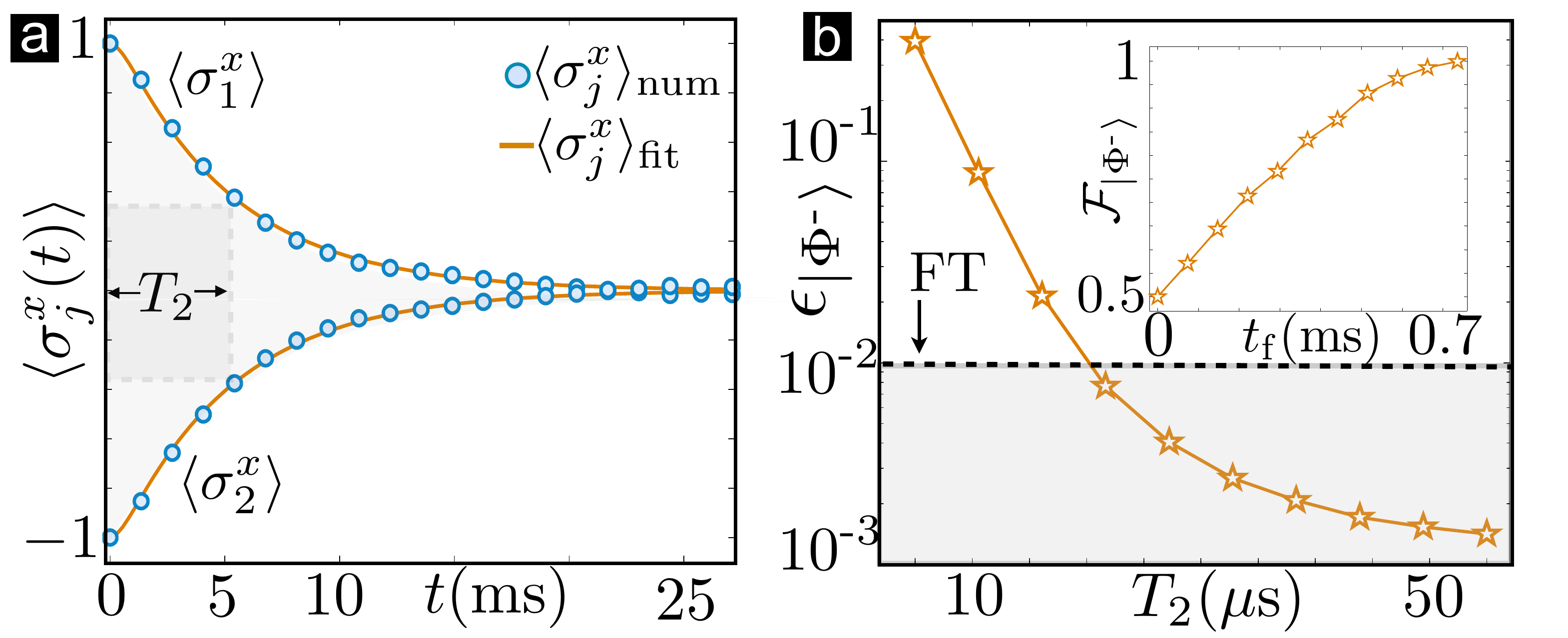}
\caption{ {\bf Resilience to magnetic-field noise:}   {\bf (a)} Exponential decay of the coherences $\langle \sigma^x(t)\rangle$ for the initial states $\ket{\psi_0}=\ket{\pm_x}$ due to the magnetic dephasing. The blue circles represent the statistical average of the numerical integration of $N=5\cdot10^3$ time evolutions, and the solid line represents the fit to an exponential decay. {\bf (b)}  Error in the generation of the Bell state $\ket{\Phi^{-}}$ from $\ket{\psi_0}=\ket{11}$ as predicted by the effective gate~\eqref{xx_model}. The spin-phonon Hamiltonian incorporates the additional magnetic-field noise, together with a strong microwave driving $\Omega_{\rm d}/2\pi=5.2$ MHz. The error is presented as a function of the dephasing $T_2$ times. In the inset, we represent the time evolution of the fidelity for the typical noise dephasing time $T_2=$5 ms.    }
\label{decoupled_gate}
\end{figure}

As an additional advantage of our scheme, we note that it also minimizes the effects of uncompensated ac-Stark shifts that may be caused by fluctuations of the laser intensities. In the derivation of the red-sideband Hamiltonian~\eqref{sideband}, we have neglected the carrier transition by imposing a weak  Rabi frequency of the laser beams. Nonetheless, the energy levels will be off-resonantly shifted due to an ac-Stark effect. In a realistic implementation of the geometric phase gates, the shifts caused by off-resonant transitions to all possible states must be  compensated by carefully selecting the laser intensities, frequencies, and polarizations~\cite{stark_shift_compensation}. However,  fluctuations of these parameters  will compromise this procedure  introducing additional noise. In clear contrast, the effects of these energy shifts in our scheme are directly cancelled by the strong driving in the same way that the energy shifts caused by magnetic-field fluctuations have been minimized.

 Let us finally comment on the effect of phase instabilities on the gate. In analogy to the conditional phase gate~\cite{phase_gate}, but in contrast to M\o lmer-S\o rensen (MS) gates~\cite{ms_gate},  our scheme does not depend on the slow drift of the  laser phases.  The second-order  process, whereby a phonon is virtually excited and then reabsorbed,  is associated to the creation and subsequent annihilation of  photons in the same pair of Raman beams,  giving rise to the insensitivity to slow changes of the phase [see Eq.~\eqref{xy_model}]. In contrast, the MS scheme involves two different pairs of Raman beams, such that the cross talk leads to the phase sensitivity. Hence, our gate~\eqref{xx_model} only relies on the phase of the microwave, which is easier to stabilize as  compared to the  phase of the two pairs of Raman  beams in the MS scheme. 
 
{\it Conclusions and outlook.--} We have introduced a scheme that merges the notion of continuous dynamical decoupling  with warm quantum gates in trapped ions. The decoupling, on top of reducing effects of external magnetic field noise, is also responsible for the gate robustness with respect to thermal fluctuations and ac-Stark shifts.  The use of continuous dynamical decoupling, as opposed to pulse sequences, yields elegant schemes which are easier to analyze theoretically and realize experimentally.  Hence,  the direction of combining quantum gates and dynamical decoupling is very promising.

Finally, we remark that our scheme can be applied to different ion species, and note that it could also be combined with a M\o lmer-S\o rensen gate to protect it from magnetic field fluctuations and ac-Stark shifts. Moreover, we emphasize that these ideas may find an application beyond the field of trapped ions. Several  quantum-information architectures also make use a bosonic data bus to couple distant qubits, whose coherence times are  limited by the environment-induced dephasing. Representative examples are trapped atoms in cavities, superconducting qubits coupled to transmission lines,   color centers in nanodiamonds coupled to nanomechanical resonators,  donor states in semiconductors coupled to phonons, or  quantum dots coupled to excitons.  The decoupling from the noise  induced  by either the environment or by the bosonic mediator can be achieved along the lines here discussed. After the submission of this work, we became aware of the interest of similar ideas for  atoms in thermal  cavities~\cite{strong_driving}, and decoherence-free states in ion traps~\cite{dressed_ions}.

{\it Acknowledgements.--}  This work was supported by the EU STREP projects HIP, PICC, the EU Integrating Projects AQUTE, QESSENCE,  by the Alexander von Humboldt Foundation, and by the DFG through QUEST.



\newpage
\newpage
\title[Short Title]{SUPPLEMENTARY MATERIAL:\\ Robust Trapped-Ion Quantum Logic Gates by Continuous Dynamical Decoupling}

\maketitle

\section*{Supplementary Material}

\begin{figure}
\centering
\includegraphics[width=\columnwidth]{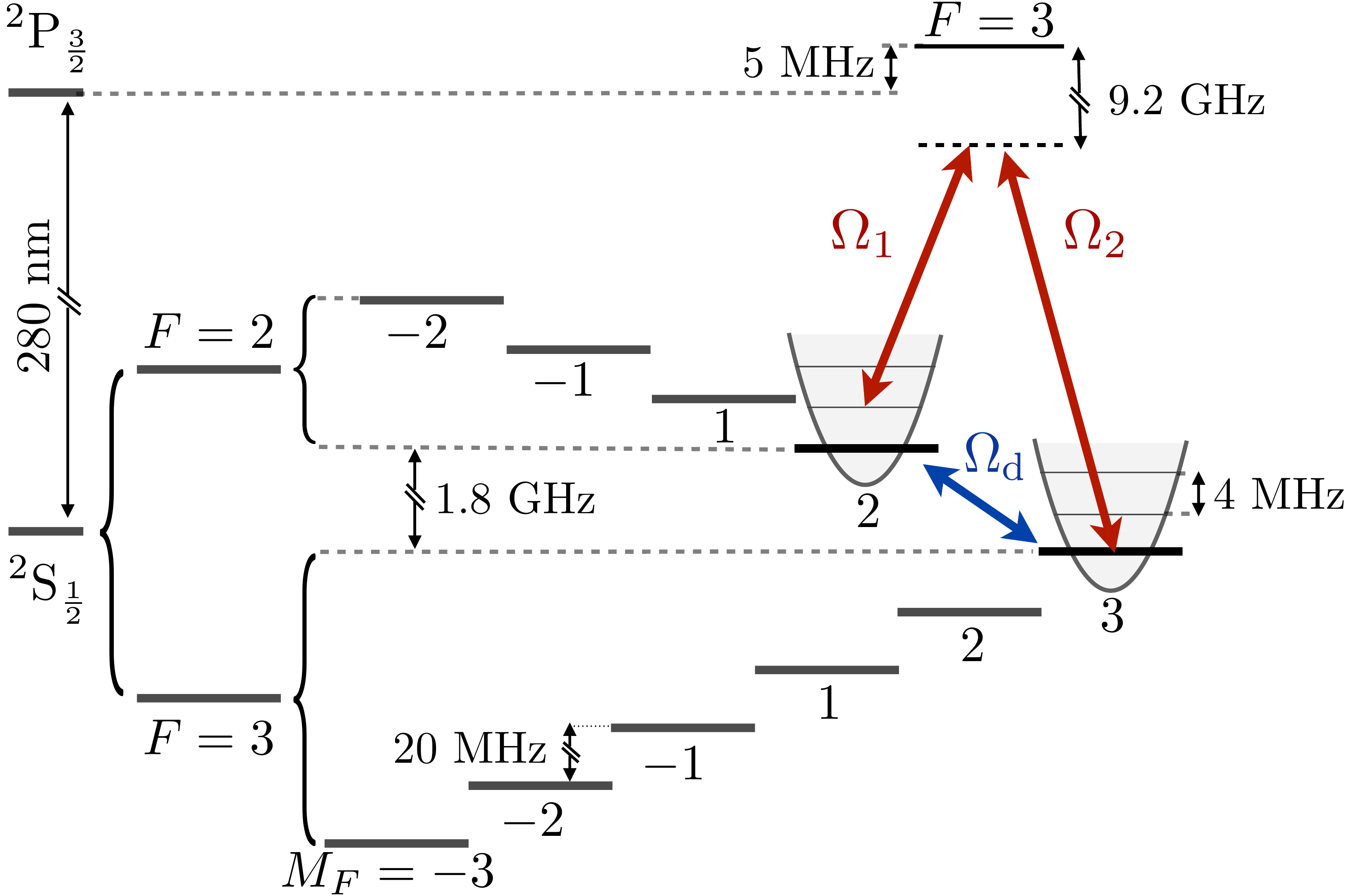}
\caption{ {\bf Hyperfine energy-level structure:}  {\bf (a)} Hyperfine structure of $^{25}{\rm Mg}^+$. The states $\ket{0}=\ket{F=3,m_{F}=3}$ and $\ket{1}=\ket{2,2}$ from the groundstate manifold $^2{\rm S}_{1/2}$ form our qubit. Two laser beams $\Omega_1,\Omega_2$ drive the red sideband via an off-resonant excited state, and a microwave $\Omega_{\rm d}$ directly couples to the transition.    }
\label{magnesium_levels}
\end{figure}

{\bf Hyperfine qubit.--} We consider the isotope of magnesium $^{25}{\rm Mg}$, which after being ionized  to $^{25}{\rm Mg}^+$  can be confined in a linear Paul trap~\cite{magnesium}. The lowest energy levels correspond to the valence electron lying in the $s$ or $p$ orbitals, which have a transition wavelength of $\lambda_{sp}\approx280$ nm. The ground-state $^2{\rm S}_{1/2}$ is split due to  hyperfine interactions into a couple of long-lived  states with total angular momentum $F=2,3$ and an energy difference of $\omega_0/2\pi\approx 1.8$ GHz [Fig.~\ref{magnesium_levels}]. Note that the transition $F=2\to3$ is electric-dipole forbidden, and the decay rate $\Gamma\approx 10^{-14}$ Hz is so slow that one can neglect spontaneous decay.  Besides, by applying an additional magnetic field, the magnetic sublevels are Zeeman split, and one can isolate two particular $\ket{F,m_F}$ states  to form our hyperfine qubit $\ket{0}=\ket{3,3},\ket{1}=\ket{2,2}$. Hence, a collection of trapped ions can be described by the pseudo-spin Hamiltonain
\begin{equation}
H=\half \sum_i\omega_0\sigma_i^z,
\end{equation}
where $\sigma_i^z=\ket{1_i}\bra{1_i}-\ket{0_i}\bra{0_i}$. These two-level systems  can be coherently manipulated by either a laser in a two-photon Raman configuration $\Omega_{1}$,$\Omega_{2}$, or a direct microwave $\Omega_{\rm d}$ [Fig.~\ref{magnesium_levels}]. Note that the external magnetic field must be strong enough so that the drivings do not excite the population of undesired Zeeman sublevels. Considering the microwave and laser Rabi frequencies employed in this work, it suffices to set $B\approx$4 mT, such that consecutive  Zeeman sublevels are split by $20$ MHz.  Alternatively, one can exploit the polarization of the microwave to select only the desired transition.

{\bf Transverse phonons.--} At low temperatures, the ion trapping forces balance the Coulomb repulsion, and the ions self-assemble in an ordered structure. The equilibrium positions are given by the minima of the confining and Coulomb potentials, and follow from the following system of equations
\begin{equation}
\tilde{z}_i^0-\sum_{j\neq i}\frac{\tilde{z}_i^0-\tilde{z}_j^0}{|\tilde{z}_i^0-\tilde{z}_j^0|^3}=0,\hspace{1ex} \tilde{z}_i^0=\frac{z_i^0}{l_z},\hspace{1ex}l_z=\left(\frac{e^2}{m\omega_z^2}\right)^{1/3}.
\end{equation}
Here, we have assumed that the transverse confinement is much tighter than the axial one, so that the ions arrange along a one-dimensional string~\cite{james}. For the two-ion crystals discussed in the main text, we have $\tilde{z}_i^0\in\{-0.63,0.63\}$. Note that for the trapping frequencies here considered $\omega_x=4\omega_z=2\pi(1 \text{MHz})$, the  ion spacing lies in the $\mu$m range.

We  now describe the properties of the transverse collective modes of the ion chain around these equilibrium positions. The small vibrations along the $x$-axis, $\Delta x_i$,  are coupled  via the Coulomb interaction, which in the harmonic approximation is 
\begin{equation}
	\label{vib}
H=\sum_{i}\left(\frac{1}{2m}{ p}_{i}^2+\frac{1}{2}m{\omega}_{x}^2\Delta x_{i}^2\hspace{-0.5ex}\right)+\frac{1}{2}\sum_{i j}\mathcal{V}_{ij}\Delta x_{i}\Delta x_{j},
\end{equation}
where $
\mathcal{V}_{ij}=m\omega_z^2(|{ \tilde{z}}^0_{ij}|^{-3}-\delta_{ij}\sum_{l\neq i}|{\tilde{z}}^0_{li}|^{-3})
$, and ${ \tilde{z}}^0_{ij}={\tilde{z}}^0_i-{ \tilde{z}}^0_j$. This quadratic  Hamiltonian can be diagonalized by means of an orthogonal transformation 
\begin{equation}
\begin{split}
\Delta x_i&=\sum_n\frac{1}{\sqrt{2m\omega_n}}\mathcal{M}_{in}(a_n^{\dagger}+a_n^{\phantom{\dagger}}),\\
 p_i&=\sum_n\ii\sqrt{\frac{m\omega_n}{2}}\mathcal{M}_{in}(a_n^{\dagger}-a_n^{\phantom{\dagger}})
\end{split}
\end{equation}
where we have set $\hbar=1$. Here,  $a_n^{\dagger},a_n^{\phantom{\dagger}}$ are the phonon creation-annihilation  operators, and $\mathcal{M}_{in}/\sqrt{2m\omega_n}$  can be interpreted as the vibrational amplitude at site ${\bf r}_i^0$ of the $n$-th normal mode characterized by the frequency $\omega_n$~\cite{feynman_lectures}. These amplitudes satisfy the following relations $\sum_{n}\mathcal{M}_{in}\mathcal{M}_{jn}=\delta_{ij}$, and $\sum_{ij}\mathcal{M}_{in}\mathcal{V}_{ij}\mathcal{M}_{jm}=\mathcal{V}_n\delta_{nm}$, where $\delta_{nm}$ is the Kronecker delta, and the normal-mode frequencies are expressed as $\omega_n=\omega_x(1+\xi\mathcal{V}_n)^{1/2}$. Here, $\xi=(\omega_z/\omega_x)^2\ll1$ quantifies the anisotropy between the axial and transverse frequencies, and also the width of the phonon branch $\omega_n\in[\omega_x(1-2\xi),\omega_x]$. For the two-ion crystals used in  the main text $\omega_x=4\omega_z$, the diagonalization of  $\mathcal{V}_{ij}$ yields   $\omega_{1}=0.968\omega_x$ for  the zig-zag mode $\mathcal{M}_{i1}=(1,-1)/\sqrt{2}$, and $\omega_{2}=\omega_x$ for the center-off mass mode $\mathcal{M}_{i2}=(1,1)/\sqrt{2}$. Taking into account these considerations, the vibrational Hamiltonian~\eqref{vib}, together with the qubit Hamiltonian, can be expressed as
\begin{equation}
\label{ho}
H_0=\frac{1}{2}\sum_i\omega_0\sigma_i^z+\sum_n\omega_n\left(a_n^{\dagger}a_n^{\phantom{\dagger}}+\half\right).
\end{equation}

\vspace{1ex}
{\bf Spin-dependent dipole forces.--} We consider the interaction of two laser beams in the Raman configuration shown as red arrows in Fig.~\ref{magnesium_levels}. These two lasers couple the two hyperfine states of our qubit  via an auxiliary excited state in the manifold $^2{\rm P}_{3/2}$. When the transitions to the excited state $F=3$ are far off-resonance~\cite{wineland_review}, this state is seldom populated and can be eliminated from the dynamics. In particular, a detuning $\Delta/2\pi\approx 9.2$GHz exceeds both the individual Rabi frequencies $\Omega_1,\Omega_2$, and the decay rate from the excited state $\Gamma$. Accordingly,   the effective coupling between the laser beam and the qubits is expressed as $H_{\rm L}=\frac{\Omega_{\rm L}}{2}\sum_i\sigma_i^+\ee^{\ii({\bf k}_{\rm L}\cdot{\bf r}_i-\omega_{\rm L}t)}+\text{H.c.}$ in the dipolar approximation. Here, we have introduced $\sigma_i^{+}=\ket{1_i}\bra{0_i}$,  the two-photon Rabi frequency  $\Omega_{\rm L}=\Omega_1^*\Omega_2/2\Delta$,  the beatnote of the laser beams $\omega_{\rm L}=\omega_1-\omega_2$,  and   the effective laser wavevector ${\bf k}_{\rm L}={\bf k}_1-{\bf k_2}$.  This wavevector is directed along the $x$-axis, and thus will only couple to the transverse phonon modes. In the interaction picture with respect to the Hamiltonian \eqref{ho}, the laser-ion interaction becomes
\begin{equation}
H_{\rm L}=\frac{\Omega_{\rm L}}{2}\sum_i\sigma_i^+\ee^{\ii \sum_n\frac{\mathcal{M}_{in}k_{\rm L}}{\sqrt{2m\omega_n}}(a_n^{\phantom{\dagger}}\ee^{-\ii\omega_nt}+a_n^{\dagger}\ee^{\ii\omega_nt})}\ee^{\ii(\omega_0-\omega_{\rm L})t}+\text{H.c.}
\end{equation}
By tuning the laser beatnote such that $\omega_{\rm L}=\omega_0-\omega_n+\delta_n$ with a detuning that fulfills $\delta_n\ll\omega_n$, it is possible to derive the red-sideband Hamiltonian $H_{\rm L}\approx H_{\rm r}(t)$ used in the main text by making a Taylor expansion in the small Lamb-Dicke (LD) parameter $\eta_n=k_{\rm L}/\sqrt{2m\omega_n}\ll1$, namely
\begin{equation}
\label{red}
H_{\rm r}(t)=\sum_{in}\mathcal{F}_{in}\sigma_i^{+}a_n^{\phantom{\dagger}}\ee^{-\ii \delta_nt}+\text{H.c.},\hspace{1ex}\mathcal{F}_{in}=\frac{\ii\Omega_{\rm L}}{2}\eta_{n}\mathcal{M}_{in}.
\end{equation} 
 Considering the trap frequency $\omega_x/2\pi=4$ MHz, and a pair of orthogonal beams such that the effective wavevector is aligned with the $x$-axis, $k_{\rm L}=\sqrt{2}(2\pi/\lambda_{sp})$, one estimates the bare LD parameter  to be $\eta=\eta_n\sqrt{\omega_n/\omega_x}\approx 0.2$ (see the Table in the main text). Note that this derivation is only valid if the Rabi frequency satisfies $\Omega_{\rm L}\ll |\omega_{\rm L}-\omega_0|$ in order to neglect  the excitation of the carrier transition in a rotating wave approximation~\cite{wineland_review}. In this work, we have considered a Raman Rabi frequency $\Omega_{\rm L}/2\pi=500$ kHz$\ll |\omega_{\rm L}-\omega_0|/2\pi\approx 3.2$MHz.

Once we have derived the red-sideband term~\eqref{red}, let us extend on its interpretation in terms of spin-dependent dipole forces outlined in the main text.  By introducing the usual Pauli matrices $\sigma_i^{\pm}=\half(\sigma_i^{x}\pm\sigma_i^y)$, the red sideband becomes
\begin{equation}
\label{red_dipole}
H_{\rm r}=\sum_{in}\hat{O}^x_{in}(t)\sigma_i^x+\hat{O}^y_{in}(t)\sigma_i^y,
\end{equation}
where we have introduced the phonon operators
\begin{equation}
\hat{O}^x_{in}(t)=\half\mathcal{F}_{in}a_n\ee^{-\ii\delta_nt}+\text{H.c.}, \hspace{1ex} \hat{O}^y_{in}(t)=\frac{\ii}{2}\mathcal{F}_{in}a_n\ee^{-\ii\delta_nt}+\text{H.c.}
\end{equation}
In order to develop an intuitive understanding of the effects of this Hamiltonian, let us initially consider a single ion subjected to one of the two non-commuting terms, say the $\sigma^x$ part. By using a Trotter decomposition for $\Delta t\to 0$, the time-evolution operator can be expressed as a concatenation of the infinitesimal unitaries $U(t+\Delta t, t)\approx D_{\ket{+_x}}(\Delta t)\ket{+_x}\bra{+_x}+D_{\ket{-_x}}(\Delta t)\ket{-_x}\bra{-_x}$, where $\ket{\pm_x}$ stand for the eigenstates of $\sigma^x$, and $D_{\ket{+_x}}(\Delta t)$ are the infinitesimal displacement operators
\begin{equation}
D_{\ket{\pm_x}}(\Delta t)=\ee^{\pm\Delta\alpha a^{\dagger}\mp\Delta\alpha^* a^{\phantom{\dagger}}},\hspace{1ex}\Delta\alpha=-\textstyle{\frac{\Omega_{\rm L}}{4}}\eta\ee^{\ii\delta t}\Delta t.
\end{equation} 
Depending on the spin state $\ket{\pm_x}$, the ion is displaced towards a particular direction in phase space. The sequential application of these displacements yields a spin-dependent closed trajectory in phase space [Fig.~\ref{dipole_forces}{\bf (a)}], where we set $\Omega_{\rm L}\in\mathbb{R}$. Considering the relation for the displacement operator $D(\alpha)D(\beta)=\ee^{\ii{\rm Im}\alpha\beta^*}D(\alpha+\beta)$, the complete time evolution will be expressed as a single displacement  with a spin-dependent phase of a geometric origin (i.e. it only depends on the area enclosed by the trajectory). This is precisely what underlies the phase gates  in ion traps~\cite{gate_reviews}. By  taking into account  the collective nature of the vibrational modes, one finds that the geometric phases depend on the spins of all ions excited by the laser beams. This allows to implement quantum logic operations  that do not depend on the thermal ion motion. 

If we consider the $\sigma^y$ term alone, a similar argument would lead to the spin-dependent trajectories displayed in Fig.~\ref{dipole_forces}{\bf (b)}, and to analogous geometric phases. However, when both $\sigma^x$ and $\sigma^y$ forces are applied simultaneously, as in the red-sideband term~\eqref{red_dipole}, the Trotter decomposition leads to phase-space trajectories that are not closed any more [Fig.~\ref{dipole_forces}{\bf (c)}]. Therefore, the geometric character of the evolution  and its independence with respect to the thermal motion of the ions are spoiled.  Below, we show explicitly how the quantum logic gates will depend upon the thermal motion of the ions.

\begin{figure}
\centering
\includegraphics[width=\columnwidth]{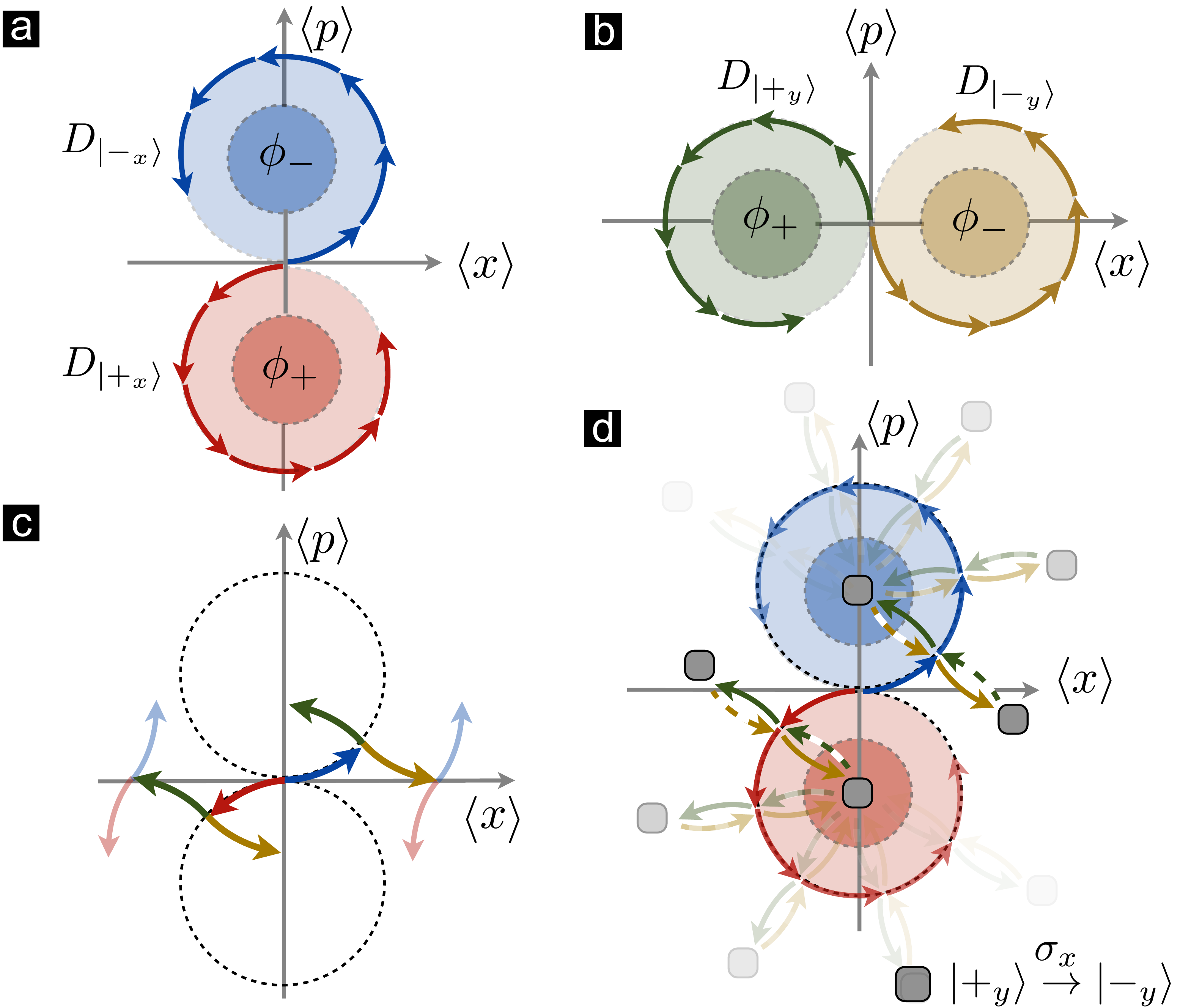}
\caption{ {\bf Spin-dependent dipole forces:}  {\bf (a)} Spin-dependent $\sigma^x$-force on a single trapped ion.  The phonons associated to spin-up states $\ket{+_x}$,$\ket{-_x}$ are displaced in phase space according to $D_{\ket{+_x}}$, $D_{\ket{-_x}}$, and form different closed paths that lead to the geometric phases $\phi_{\pm}$. {\bf (b)} Spin-dependent $\sigma^y$-force. {\bf (c)} Trotterization of the combined $\sigma^x$ and $\sigma^y$ forces. The $\sigma^x$-displacement $D_{\ket{+_x}}(\Delta t)$ shall be followed by the two possible $\sigma^y$ displacements $D_{\ket{\pm_y}}(\Delta t)$ since $\ket{+_x}\propto (\ket{+_y}+\ii\ket{-_y}$). Hence, the phase-space trajectory is not generally closed. {\bf (d)} Schematic spin-echo refocusing of the $\sigma^y$-force. By applying a $\pi$-pulse  $X^{\pi}_i=\sigma^x_i$ (grey box), $\ket{\pm_y}\to\ket{\mp_y}$, the  displacements $D_{\ket{\pm}_y}$ are reversed (dotted arrows), such that the trajectory is refocused and the closed paths yield again a geometric phase.     }
\label{dipole_forces}
\end{figure}

\begin{figure}
\centering
\includegraphics[width=0.95\columnwidth]{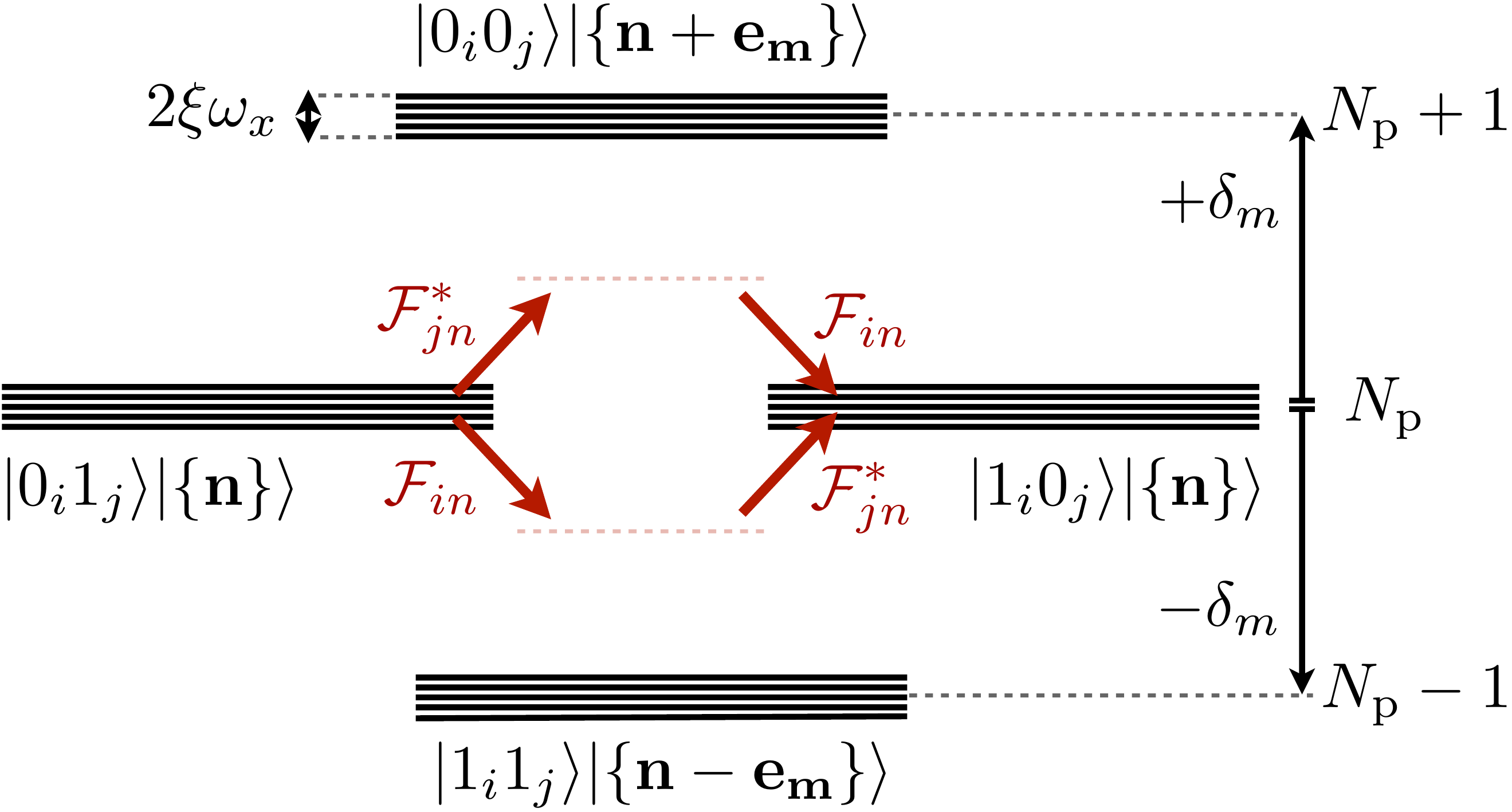}
\caption{ {\bf Effective flip-flop interactions:} In the limit of large detuning $|\mathcal{F}_{in}|\ll\delta_{n}$, the energy spectrum of the Hamiltonian~\eqref{ad} is clustered in manifolds characterized by the total number of vibrational excitations $\ket{\{\bf n\}}$, such that $\sum_m n_m=N_{\rm p}$. The energy width of these manifolds is bounded by $|\delta_n-\delta_m |=2\xi\omega_x$, where $\xi=(\omega_z/\omega_x)^2\ll1$.  The second order processes (red arrows) where a phonon is created and then reabsorbed (or viceversa), give rise to a flip-flop interaction where the qubit state is exchanged between two distant ions.}
\label{flip_flop}
\end{figure}

\vspace{1ex}
{\bf Interactions by virtual phonon exchange.--} 
 The full Hamiltonian $H_0+H_{\rm r}$ can be expressed in a picture where the phonon frequency is replaced by the detuning with respect to the particular sideband, namely
\begin{equation}
\label{ad}
H=\sum_n\delta_na_n^{\dagger}a^{\phantom{\dagger}}_n+\sum_{in}(\mathcal{F}_{in}\sigma_i^+a^{\phantom{\dagger}}_n+\mathcal{F}_{in}^*\sigma_i^-a^{\dagger}_n).
\end{equation}
This expression is now amenable to perform the adiabatic elimination of the phonons (i.e. quasi-degenerate perturbation theory~\cite{cohen_atom_photon}) , and obtain the effective coupling between the qubits. In the limit of large detuning, $|\mathcal{F}_{in}|\ll\delta_{n}$, only virtual phonon excitations can take place [Fig.~\ref{flip_flop}].  There are two possible paths for the virtual phonon exchange between two distant ions: {\it i)} Either the system virtually populates the manifold with  one extra phonon, or {\it ii)} it goes through a lower-energy manifold with one phonon less. Since these two processes have an opposite detuning, their amplitudes cancel $(\mathcal{F}_{in}\mathcal{F}_{jm}^*)(1/\delta_n+1/\delta_m)-(\mathcal{F}^*_{in}\mathcal{F}_{jm})(1/\delta_n+1/\delta_m)=0$, except when the exchanged phonon belongs to the same mode $n=m$. In such a case, one has to take into account the bosonic nature of the phonons $a^{\phantom{\dagger}}_na_n^{\dagger}=1+a_n^{\dagger}a^{\phantom{\dagger}}_n$, which spoils the interference between both exchange paths, and leads to an effective Hamiltonian where the phonons have been eliminated
\begin{equation}
\label{xy}
H_{\rm eff}=\sum_{ij} J^{\rm eff}_{ij}\sigma_i^+\sigma_j^-,\hspace{1ex} J^{\rm eff}_{ij}=\sum_n\frac{-1}{\delta_n}\mathcal{F}_{in}\mathcal{F}_{jn}^*.
\end{equation}
At this point, we should note that there is a missing process in the above argument, namely that the virtually-excited phonon can be reabsorbed by the same ion. In this case, it does not lead to a phonon-mediated interaction. Due to the algebraic properties of the Pauli matrices, the above cancellation between the two paths does not take place in this case, and one obtains an ac-Stark shift that depends on the phonons, namely 
\begin{equation}
H_{\rm res}=\sum_{inm}B_{inm}a_m^{\dagger}a_n^{\phantom{\dagger}}\sigma_i^z, \hspace{1ex}B_{inm}=-\half\mathcal{F}_{in}\mathcal{F}_{im}^*(\frac{1}{\delta_n}+\frac{1}{\delta_m}).
\end{equation}
This term   has a dramatic effect on the quantum logic gates since it couples the spins to the phonons with a strength similar to that of the desired gate. Hence, the spin dynamics becomes sensitive to the thermal motion of the ions, and the geometric character of the phase gates is lost. We note that this term may become a gadget, rather than a limitation, in the  quantum simulation of Anderson localization of  the vibrational excitations in an ion chain~\cite{xy_model}.

\vspace{1ex}
{\bf Polaron transformation for a strong driving.--}
So far, we have considered the phonon-mediated qubit-qubit interactions due to the red sideband. We now describe the effects of a strong driving of the carrier transition
\begin{equation}
\label{driving}
H_{\rm d}(t)=\half\sum_i{\Omega_{\rm d}}\ee^{\ii\phi_{\rm d}}\sigma_i^{+}(\ee^{\ii(\omega_0-\omega_{\rm d})t}+\ee^{\ii(\omega_0+\omega_{\rm d})t})+\text{H.c.},
\end{equation}
where  $\Omega_{\rm d}$ is the driving Rabi frequency, and $\phi_{\rm d}$ its phase, and $\omega_{\rm d}$ its frequency. Let us recall that for our qubit choice, this driving can be realized with microwaves.
We consider a resonant microwave  $\omega_{\rm d}=\omega_0$ with $\Omega_{\rm d}\ll\omega_{\rm d}$, so that one can neglect the counter-rotating terms in Eq.~\eqref{driving}. In the picture introduced above~\eqref{ad}, the driven Hamiltonian becomes
\begin{equation}
H=\sum_n\delta_na_n^{\dagger}a^{\phantom{\dagger}}_n+\frac{\Omega_{\rm d}}{2}\sum_i(\ee^{\ii\phi_{\rm d}}\sigma_i^++\text{H.c.})+\sum_{in}(\mathcal{F}_{in}\sigma_i^+a^{\phantom{\dagger}}_n+\text{H.c.}).
\end{equation}
For the sake of simplicity, we set $\phi_{\rm d}=0$ since it does not change the essence of the decoupling mechanism described below. By moving onto the dressed-state basis, $\ket{\pm_x}_i=(\ket{1_i}\pm\ket{0_i})/\sqrt{2}$, the red-sideband term  becomes
\begin{equation}
\label{red_dressed}
\begin{split}
H_{\rm r}(t)&=\sum_{in}\half\mathcal{F}_{in}\big(\ket{-_x}_i\bra{+_x}\ee^{-\ii\Omega_{\rm d}t}-\ket{+_x}_i\bra{-_x}\ee^{+\ii\Omega_{\rm d}t}\big)a^{\phantom{\dagger}}_n\ee^{-\ii\delta_nt}\\
&+\sum_{in}\half\mathcal{F}_{in}\big(\ket{+_x}_i\bra{+_x}-\ket{-_x}_i\bra{-_x}\big)a^{\phantom{\dagger}}_n\ee^{-\ii\delta_nt}+\text{H.c.}\
\end{split}
\end{equation}
From this expression, one readily observes that the terms involving transitions between the dressed eigenstates rotate very fast for $\Omega_{\rm d}\gg \delta_n$, and their contribution to the effective interactions will be negligible. In order to give a stronger weight to the terms diagonal in the dressed-state basis, we perform the spin-dependent displacement $a_n\to a_n-\sum_i\mathcal{F}_{in}^*\sigma_i^{x}/2\delta_n$. This canonical transformation is formalized in terms of a Lang-Firsov-type polaron transformation~\cite{lang_firsov} as follows
\begin{equation}
U=\ee^{S},\hspace{1ex}S=\sum_{in}{\frac{\mathcal{F}_{in}^*}{2\delta_n}}\sigma_i^xa_n^{\dagger}-\text{H.c.},
\end{equation}
which has also been used in the context of trapped ions~\cite{microwave_gates_wunderlich,ising_porras}. This unitary offers an alternative mechanism to obtain the spin interactions by virtual phonon exchange, and also allow us to calculate all the residual spin-phonon couplings to any desired order of the small parameter $|\mathcal{F}_{jn}|/\delta_n\ll1$. In particular, considering the algebraic properties of the spins and phonons,  it transforms the relevant operators as follows
\begin{equation}
\begin{split}
\label{transformation}
Ua_nU^{\dagger}&=a_n-\sum_j\frac{\mathcal{F}_{jn}^*}{2\delta_n}\sigma_j^x,\\
U\sigma_i^yU^{\dagger}&=\text{cosh}(\hat{\Theta}_i)\sigma_i^y-\ii\sinh(\hat{\Theta}_i)\sigma_i^z,\\
\end{split}
\end{equation}
where $\hat{\Theta}_i=\sum_m\mathcal{F}_{im}a_m^{\phantom{\dagger}}/2\delta_m-\text{H.c.}$ Therefore, it displaces the phonon operators, whereas it rotates the spins around the $x$-axis. These transformations yield the effective Hamiltonian 
\begin{equation}
\label{xx_model}
\tilde{H}_{\rm eff}=\sum_{ij} \tilde{J}^{\rm eff}_{ij}\sigma^x_i\sigma^x_j+\half\sum_i\Omega_{\rm d}\sigma^x_i,\hspace{1ex}\tilde{J}^{\rm eff}_{ij}=\fourth J^{\rm eff}_{ij},
\end{equation}
 and the residual spin-phonon coupling
\begin{equation}
\tilde{H}_{\rm res}=\sum_{in}\frac{\ii}{2}\left(\mathcal{F}_{in}a_n^{\phantom{\dagger}}-\mathcal{F}_{in}^*a_n^{\dagger}\right)\left(\text{cosh}\hat{\Theta}_i\sigma_i^y-i\text{sinh}\hat{\Theta}_i\sigma_i^z\right).
\end{equation}
To any order of perturbation theory, the residual spin-phonon couplings only involves terms that try to induce transitions between the dressed eigenstates, namely $\sigma_i^y=\ii\ket{+_i}\bra{-_i}+\text{H.c.}$, and $\sigma_i^z=\ket{+_i}\bra{-_i}+\text{H.c.}$. Therefore, in the limit of strong microwave driving $\Omega_{\rm d}\gg\Omega_{\rm L}$,  the residual spin-phonon term  becomes rapidly rotating and can be neglected in a rotating wave approximation $\tilde{H}_{\rm res}\approx 0$. For this approximation to hold, we have to make sure that  there is no resonance with a process that does not conserve the number of phonons. For the parameters of interest to our setup, it suffices to consider that $\Omega_{\rm d}\gg2\delta_n$ to avoid such processes, and thus minimize the effects of the residual spin-phonon coupling. 

By refocusing the fast rotations induced by the resonant microwave (as considered in the main text), the effective Hamiltonian becomes $\tilde{H}_{\rm eff}=\sum_{ij} \tilde{J}^{\rm eff}_{ij}\sigma^x_i\sigma^x_j$, which is responsible of a geometric phase gate insensitive to the thermal motion of the ions. After the time $t_{\rm f}=\pi/(8\tilde{J}^{\rm eff}_{ij})$, 
this gate $U_{\rm eff}(t_{\rm f})=(1-\ii\sigma_i^x\sigma_j^x)/\sqrt{2}$ generates the four entangled Bell states
\begin{equation}
\label{phase_gate}
\begin{split}
\ket{0_i0_j}\to\textstyle{\frac{1}{\sqrt{2}}}\ket{0_i0_j}-\frac{\ii}{\sqrt{2}}\ket{1_i1_j},\\
\ket{0_i1_j}\to\textstyle{{\frac{1}{\sqrt{2}}}\ket{0_i1_j}-\frac{\ii}{\sqrt{2}}\ket{1_i0_j}},\\
\ket{1_i0_j}\to\textstyle{{\frac{1}{\sqrt{2}}}\ket{1_i0_j}-\frac{\ii}{\sqrt{2}}\ket{0_i1_j}},\\
\ket{1_i1_j}\to\textstyle{{\frac{1}{\sqrt{2}}}\ket{1_i1_j}-\frac{\ii}{\sqrt{2}}\ket{0_i0_j}},
\end{split}
\end{equation}
and together with single-qubit rotations it constitutes a universal set for quantum computation. At this point, it is worth going back to the intuitive interpretation of the spin-dependent forces, and the associated geometric phase gates [Fig.~\ref{dipole_forces}]. The effect of the strong driving can be understood as a continuous version of the $\sigma^x$-pulses that refocuses the displacement caused by the  $\sigma^y$ force [Fig.~\ref{dipole_forces}{\bf (d)}]. With this intuitive picture, it is clear that we have recovered the geometric character of the gate, and  the robustness with respect to thermal motion.

\vspace{1ex}
{\bf Drifts of the laser and microwave phases.--} We clarify the role of  the phases of the  red-sideband and carrier excitations.

 In the red-sideband term~\eqref{red}, we have assumed a vanishing phase $\Omega_{\rm L}\in\mathbb{R}$. Note that the effective interaction in Eq.~\eqref{xy} is not modified by considering a different phase $\Omega_{\rm L}\to\Omega_{\rm L}\ee^{\ii\varphi_{\rm L}}$, or equivalently $\mathcal{F}_{in}\to\mathcal{F}_{in}\ee^{\ii\varphi_{\rm L}}$. Intuitively, one may consider that this phase only affects the trajectory of the displaced phonons, but not the geometric phase~\cite{gate_reviews}. Formally, this is a consequence of the perturbative  process where phonon excitations are virtually created and annihilated, such that   the  qubit-qubit couplings only depend on terms like $\mathcal{F}_{in}\mathcal{F}_{jn}^{*}$, which are independent of the laser phases. According to the Raman process [Fig.~\ref{magnesium_levels}], this type of terms follow from the absorption and emission of  a photon in each of the two laser beams,  thus canceling the dependence on the laser phase. Note that the same insensitivity occurs in the presence of a strong driving~\eqref{xx_model}. This is a clear advantage with respect to M$\o$lmer-S$\o$rensen gates~\cite{molmer_sorensen_gate}, since those depend on the phases of two different sidebands that follow from a bichromatic laser beam, and are thus subjected to fluctuations in the optical path length.
 
We have also assumed a vanishing phase of the carrier interaction~\eqref{driving}, and argued that the essence of the decoupling mechanism is not altered. Let us consider the effects of a non-vanishing phase $\phi_{\rm d}$. In this case,  the dressed-state basis must be modified to $\ket{\pm_{\rm d}}_i=(\ket{1_i}\pm\ee^{-\ii\phi_{\rm d}}\ket{0_i})/\sqrt{2}$, and thus the red-sideband excitation~\eqref{red_dressed} in that basis becomes
\begin{equation}
\begin{split}
H_{\rm r}(t)&\!=\!\sum_{in}\textstyle{\frac{\ee^{-\ii\phi_{\rm d}}}{2}}\mathcal{F}_{in}\big(\ket{-_d}_i\bra{+_d}\ee^{-\ii\Omega_{\rm d}t}\!-\!\ket{+_d}_i\bra{-_d}\ee^{+\ii\Omega_{\rm d}t}\big)a^{\phantom{\dagger}}_n\ee^{-\ii\delta_nt}\\
&+\sum_{in}\textstyle{\frac{\ee^{-\ii\phi_{\rm d}}}{2}}\mathcal{F}_{in}\big(\ket{+_d}_i\bra{+_d}-\ket{-_d}_i\bra{-_d}\big)a^{\phantom{\dagger}}_n\ee^{-\ii\delta_nt}+\text{H.c.}
\end{split}
\end{equation}
Note that we can still apply the same rotating wave approximation when $\Omega_{\rm d}\gg\delta_n$, and thus neglect some of the terms in the above expression. This encourages a modification of the polaron-type transformation $a_n\to a_n-\sum_i\mathcal{F}_{in}^*\ee^{\ii\phi_{\rm d}}\sigma_i^{\rm d}/2\delta_n$, where we have introduced the dressed spin operator $\sigma_i^{\rm d}=\ee^{\ii\phi_{\rm d}}\sigma_i^++\ee^{-\ii\phi_{\rm d}}\sigma_i^-$. By the same argument used above, we see that the phase appearing in the polaron transformation does not play any role. Hence, we would obtain an the same effective Hamiltonian as in Eq.~\eqref{xx_model}, but changing $\sigma_i^x\to\sigma_i^{\rm d}$. We have thus arrived at the final conclusion that the effective Hamiltonian does only depend on the phase of the carrier excitation, but not on the phase of the red sideband. This is a clear advantage for our qubit choice, since the carrier phase comes from a microwave source, which is more stable than the laser beams leading to the sideband.

\vspace{1ex}
{\bf Magnetic noise model.--}  The internal level structure of trapped ions can be perturbed by uncontrolled external electric and magnetic fields. Since the  Stark shifts are typically small~\cite{wineland_review}, one is only  concerned with magnetic-field fluctuations of the resonance frequency $\omega_0\to\omega_0+\partial_{B}\omega_0(B-B_0)+\half\partial_{B^2}\omega_0(B-B_0)^2$. For clock states, either there is no such magnetic-field dependence (e.g. $\ket{0_i},\ket{1_i}$ have both zero magnetic moment), or the linear Zeeman shift vanishes at a certain magnetic field. In such cases, the internal state coherence times can reach even minutes. On the other hand, when none of the above conditions holds, one refers to magnetic-field sensitive states whose typical coherence times are reduced by magnetic-field fluctuations to $T_2\sim$1-10ms. The fluctuation of the resonance frequency introduces a term in the Hamiltonian $H_{\rm n}=\half\sum_iF(t)\sigma_i^z$, where $F(t)=-g\mu_{\rm B}B(t)$ fluctuates,  $\mu_{\rm B}$ is the Bohr magneton, and $g$ the hyperfine $g$-factor. Low-frequency fluctuations (i.e. fluctuations that only take place between different experimental runs) can be easily prevented by using spin-echo sequences~\cite{spin_echo}. On the other hand, the effects of fast-frequency fluctuations cannot be refocused unless complicated dynamical-decoupling sequences are used. 

We are concerned with this  type of fast fluctuations, which lead to an exponential decay of the coherences $\langle \sigma_i^x(t)\rangle$ as measured by Ramsey interferometry~\cite{wineland_review}. In order to reproduce such an exponential decay, we use a paradigmatic model of noise where $F(t)$ corresponds to a stationary, Markovian, and gaussian random process. Such a stochastic process is known as a Ornstein-Uhlenbeck process~\cite{random_processes,ou_process}, and is characterized by the following Langevin equation 
\begin{equation}
\frac{{\rm d}F(t)}{{\rm d}t}=-\frac{F(t)}{\tau}+\sqrt{c}\Gamma(t),
\end{equation}  
where $c$  is the diffusion constant, $\tau$ the correlation time, and  $\Gamma(t)$ is a gaussian white noise that fulfills $\langle \Gamma\rangle_{\rm st}=0$, $\langle \Gamma(t)\Gamma(0)\rangle_{\rm st}=\delta(t)$. This particular stochastic differential equation can be integrated exactly yielding a gaussian random process with the following mean
and variance
\begin{equation}
\begin{split}
\langle F\rangle_{\rm st}&=F(t_0)\ee^{-(t-t_0)/\tau},\\
 \langle F^2\rangle_{\rm st}- \langle F\rangle_{\rm st}^2&=\textstyle{\frac{c\tau}{2}}(1-\ee^{-2(t-t_0)/\tau}),
\end{split}
\end{equation}
which show that the correlation time $\tau$ sets the time scale over which the process relaxes to the asymptotic values. Besides, the autocorrelation function  $\langle F(t)F(0)\rangle_{\rm st}=\frac{c\tau}{2}\ee^{-t/\tau}$ shows that $\tau$ also sets the time scale such that the values of process are correlated or not, where we have assumed a vanishing mean as customary. Of primary importance to the numerical simulations is the update formula
\begin{equation}
\label{noise}
F(t_2)=F(t_1)\ee^{-\frac{\delta t}{\tau}}+\big[\textstyle{\frac{c\tau}{2}}(1-\ee^{-\frac{2\delta t}{\tau}})\big]^{\frac{1}{2}}n,
\end{equation}
where $n$ is a unit gaussian random variable: This formula is valid for an arbitrary discretization $t_2=t_1+\delta t$,~\cite{ou_process}, and is thus ideally suited to for the numerical integration.

The single-qubit  coherences are affected by  this type of noise. Considering the initial state $\ket{\psi_0}=\ket{\pm_x}$, one obtains 
\begin{equation}
\langle\sigma^x(t)\rangle=\pm\ee^{-\half\langle\varphi^2(t)\rangle_{\rm st}},
\end{equation}
 where $\varphi(t)=\int {\rm d}t'F(t')$ is the random process given by the time integral of $F(t)$, and its autocorrelation function is
 \begin{equation}
\langle\varphi^2(t)\rangle_{\rm st}=c\tau^2\left(t-\tau\left(\frac{3}{2}-2\ee^{-t/\tau}+\half\ee^{-2t/\tau}\right)\right).
\end{equation}
 In the limit of short correlation times $\tau\ll t$, one obtains an exponential damping  with a decay time $T_2=2/c\tau^2$.

Let us finally comment on the modification of the Lang-Firsov transformation in the presence  of the magnetic noise~\eqref{noise}, which would lead to extra residual spin-phonon couplings. 
However, these contributions are negligible for the regime of interest  $T_2\sim 1$-$10$ ms. This can be understood by considering the variance at long times, which leads to $\sqrt{\langle F^2\rangle_{\rm st}}=\sqrt{1/\tau T_2}$. Since the magnetic-field noise in ion traps corresponds to the limit $\tau\ll t$, we have considered $\tau=0.1 T_2$ throughout the text, which leads to a typical noise strength in the 0.1-1 kHz regime. Accordingly, $\Omega_{\rm d}\gg\sqrt{\langle F^2\rangle_{\rm st}}$, and the residual spin-phonon couplings that come from the Lang-Firsov transformation can be  neglected. 

\vspace{1ex}
{\bf Fidelity of the quantum gate.--} In the main text of this manuscript, we have quantified the efficiency of the quantum gate for particular entangled Bell states. However, it is also desirable to evaluate how close is the real time-evolution of the system to the desired  unitary $U_{\rm eff}=\ee^{-\ii\frac{\pi}{4}\sigma_i^x\sigma_j^x}$. This can be quantified by the following fidelity
\begin{equation}
\label{gate_fidelity}
\mathcal{F}(U_{\rm eff},\mathcal{E})=\int d\psi_{\rm s}\bra{\psi_{\rm s}}U_{\rm eff}^{\dagger}\mathcal{E}(\ket{\psi_{\rm s}}\bra{\psi_{\rm s}})U_{\rm eff}\ket{\psi_{\rm s}},
\end{equation}
where one integrates over the whole Hilbert space of two-qubit states $\ket{\psi_{\rm s}}$, and $\mathcal{E}$ is the quantum channel that describes the real time-evolution of the trapped-ion setup
\begin{equation}
\mathcal{E}(\ket{\psi_{\rm s}}\bra{\psi_{\rm s}})={\rm Tr}_{\rm ph}\{U(t_{\rm f})(\rho_{\rm th}\otimes\ket{\psi_{\rm s}}\bra{\psi_{\rm s}})U(t_{\rm f})^{\dagger}\},
\end{equation}
where  $U(t_{\rm f})$ describes the time-evolution of the complete spin-phonon system including the stochastic average over the magnetic field noise, and $\rho_{\rm th}$ corresponds to the thermal Gibbs state for the phonons. The  integral over the state space in Eq.~\eqref{gate_fidelity} makes the computation of the quantum channel fidelity rather inefficient, specially if one considers the additional random sampling over the magnetic field noise. An alternative to this method is the evaluation of the so-called entanglement fidelity for the channel~\cite{ent_fidelity} , which can be expressed as follows 
\begin{equation}
\label{ent_fidelity}
\mathcal{F}_{\rm e}(U_{\rm eff},\mathcal{E})=\bra{\phi_{\rm m}}\mathbb{I}_d\otimes U_{\rm eff}^{\dagger}\mathcal{E}(\ket{\phi_{\rm m}}\bra{\phi_{\rm m}}) \mathbb{I}_d\otimes U_{\rm eff}\ket{\phi_{\rm m}},
\end{equation}
where $\ket{\phi_{\rm m}}=\sum_{\alpha=1}^d\ket{\alpha}\otimes\ket{\alpha}/\sqrt{d}$ is the maximally entangled state between the physical two-qubit system, and an ancillary copy of it. Hence, $\ket{\alpha}\in\{\ket{0_i,0_j},\ket{0_i,1_j},\ket{1_i,0_j},\ket{1_i,1_j}\}$, and $d=4$. In the expression of the entanglement fidelity, the quantum channel does not act on the ancillary qubits, namely
\begin{equation}
\mathcal{E}(\ket{\phi_{\rm m}}\bra{\phi_{\rm m}})={\rm Tr}_{\rm ph}\{\mathbb{I}_{d}\otimes U(t_{\rm f})(\rho_{\rm th}\otimes\ket{\phi_{\rm m}}\bra{\phi_{\rm m}})\mathbb{I}_{d}\otimes U(t_{\rm f})^{\dagger}\}.
\end{equation}
The crucial point is that this entanglement fidelity is related to the quantum channel fidelity~\cite{fidelity} by the following expression
\begin{equation}
\label{fidelity_relation}
\mathcal{F}(U_{\rm eff},\mathcal{E})=\frac{d\mathcal{F}_{\rm e}(U_{\rm eff},\mathcal{E})+1}{d+1}.
\end{equation}
Therefore, it suffices to calculate the entanglement fidelity~\eqref{ent_fidelity}, which turns out to be computationally less demanding in the present setup, and then infer the value of the quantum channel fidelity~\eqref{gate_fidelity} via Eq.~\eqref{fidelity_relation}. 

\begin{figure}
\centering
\includegraphics[width=0.95\columnwidth]{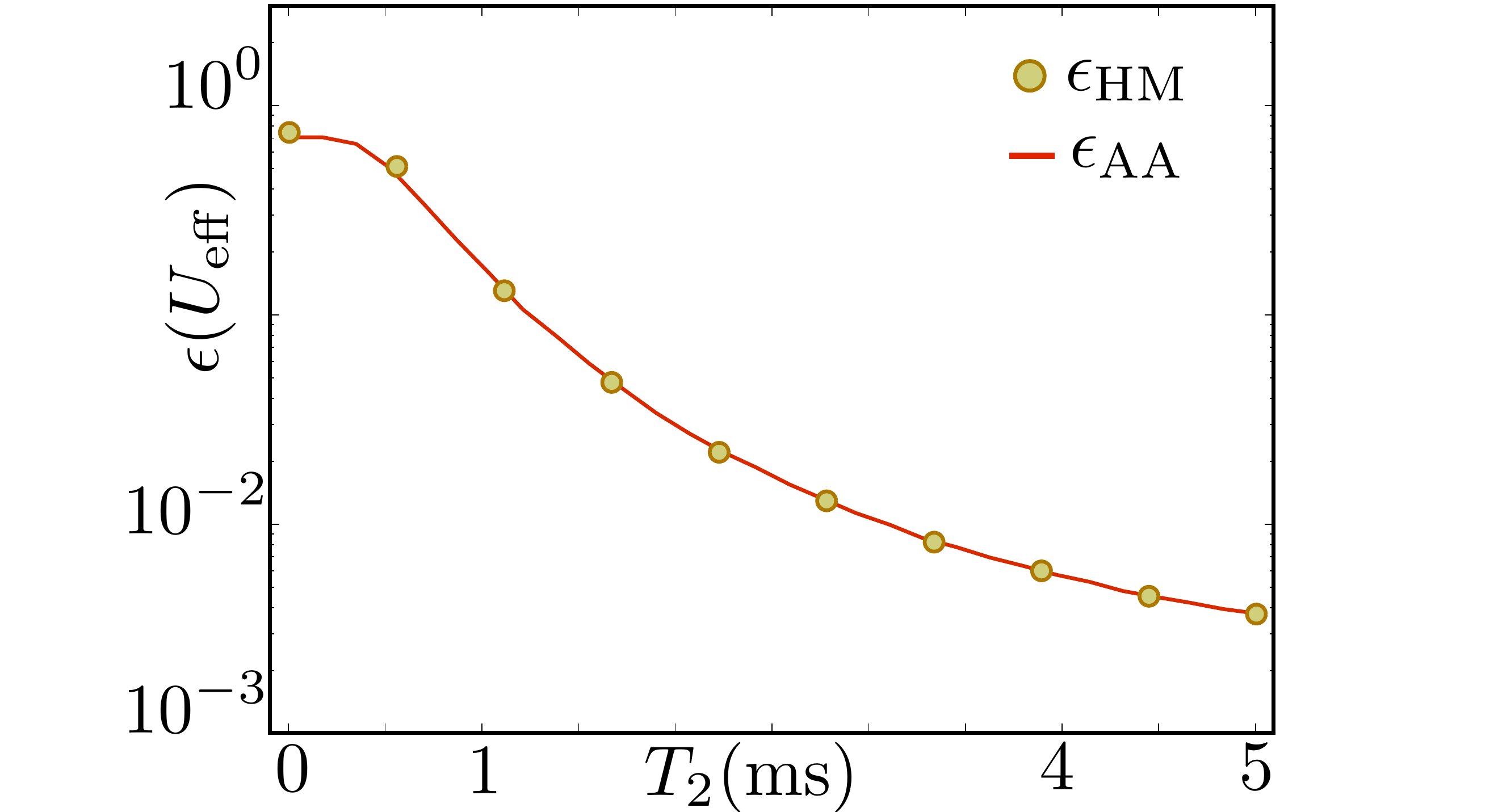}
\caption{ {\bf Quantum channel error:} Error $\epsilon(U_{\rm eff})$ of the quantum channel $\mathcal{E}$ for the time-evolution of the two-ion system with respect to the desired quantum gate  $U_{\rm eff}$. The error is estimated according to the discretized version $\epsilon_{\rm HM}$ of Eq.~\eqref{gate_fidelity} (yellow dots), and compared to the ancillary-assisted estimate $\epsilon_{\rm AA}$(red solid line) in~\eqref{fidelity_relation}.}
\label{channel_fidelity}
\end{figure}

In Fig.~\ref{channel_fidelity}, we present our results for this ancillary-assisted estimation of the  quantum gate error $\epsilon_{\rm AA}=1-\mathcal{F}_{\rm AA}$ (red line), and compare it to a discretized version of the Haar measure in~\eqref{gate_fidelity}, which we have denoted as $\epsilon_{\rm HM}=1-\mathcal{F}_{\rm HM}$ (yellow dots). To perform the integral over state space, we sample randomly over $N_{\rm s}=10^3$ different initial states, and average the corresponding fidelities. Let us also note that we have considered $N_{\rm d}=10$ different dephasing times $T_2\in[0,5]$ ms,  each of which requires a statistical average over $N_{\rm m}=5\cdot 10^3$ different histories of the stochastic process describing the magnetic-field  noise. Accordingly, the estimation of $\epsilon_{\rm HM}$ requires the numerical integration of $N=5\cdot 10^7$ evolutions of the spin-phonon system, which is more demanding than the ancillary-based method. From the results  in Fig.~\ref{channel_fidelity}, we conclude that the quantum channel error  approaches the fault-tolerance region even in the presence of the magnetic-field noise.  

\vspace{1ex}
{\bf Comparison to state-of-the-art gate implementations.--} Finally, let us compare our scheme for robust quantum gates with other existing protocols. We place a special emphasis on the state-of-the-art experimental implementations where the natural noise sources limiting the fidelities have been discussed (see~\cite{gate_reviews} and references therein).
 
 Let us first consider the implementations of the Cirac-Zoller scheme for two-ion quantum gates with the highest fidelities $\mathcal{F}\approx0.92$ for gate times $t_{\rm f}\approx 0.5$ ms~\cite{cz_gates}. These fidelities were 
limited by two error sources, namely,  off-resonant contributions to the carrier transition, and fluctuating laser frequencies and  magnetic fields shifting the resonance frequency. Additionally, these experiments require single-ion addressing and a perfect laser cooling to the motional ground-state, which can be hard to extend to larger ion registers. As shown in the main text, our scheme overcomes all these problems. The carrier is continuously driven so that  the above off-resonant contribution plays no significant role (note, however, that we must still be careful with the off-resonant contributions to other Zeeman sublevels). The fluctuations of the  resonance frequency are cured by the continuous decoupling mechanism. Besides, single-ion addressing during the gate is not required, and we have shown that there is no necessity of perfect ground-state cooling, while  potentially attaining higher fidelities $\mathcal{F}\approx0.99-0.999$ with similar speeds $t_{\rm f}\approx0.7$ ms. 

Let us now compare our scheme to the so-called $\sigma^z$ geometric  phase gates~\cite{didi_gates}, which achieved a fidelity $\mathcal{F}\approx0.97$ for two-qubit gates with speeds $t_{\rm f}\approx 0.04$ ms. Our gate scheme shares some of the advantages of this gate, such as the robustness to thermal fluctuations of the ion motion, the absence of single-ion addressability, and the resilience to off-resonant contributions to the carrier transition.   It is precisely the latter which allows the $\sigma^z$-gate to attain higher speeds as compared to the above Cirac-Zoller gates. In fact, these speeds are required to achieve such high fidelities in the presence of the errors due to due to drifts of the laser frequencies and fluctuating magnetic fields. Besides, the photon scattering due to the Raman beam arrangement introduces additional decoherence. Our scheme overcomes the fluctuations in the resonance frequency by means of the continuous decoupling, which allows us to boost the coherence times far beyond the millisecond regime,  reaching higher fidelities $\mathcal{F}\approx0.99-0.999$ for slower gates $t_{\rm f}\approx0.7$ ms. In fact, the dynamical decoupling allows still slower gates so that we may use a lower Rabi frequency for the Raman beams, or alternatively a  larger detuning with respect to the auxiliary dipole-allowed transition, thus reducing further the noise by photon scattering~\cite{plenio_sp}. Let us finally note that a source of dephasing noise common to both gate schemes is that of fluctuating Rabi frequencies of the spin-phonon couplings. This problem can be overcome by stabilizing the laser intensities on long time scales.

We now compare our proposal to  the M\o lmer-S\o rensen scheme for hyperfine qubits~\cite{ms_gates,ms_clock}, also known as the $\sigma^{\phi}$ geometric  phase gate. The best fidelities achieved  $\mathcal{F}\approx0.89$ for gate speeds $t_{\rm f}\approx 0.08$ ms were limited by fluctuating ac-Stark shifts and photon scattering from the auxiliary excited state~\cite{ms_clock}. Our scheme solves the fluctuating ac-Stark shifts by the dynamical decoupling, and may minimize the photon scattering at the expense of using slower gates (note that the gates are still much faster than the coherence time enhanced  via the  decoupling mechanism). Hence, it can reach higher fidelities $\mathcal{F}\approx0.99-0.999$ while preserving the advantages of the $\sigma^{\phi}$-gate, which is insensitive to thermal fluctuations~\cite{ms_gates,ms_clock},  uses a laser beam configuration that makes it resilient to slow drifts of the laser phases~\cite{ms_clock}, and overcomes magnetic field fluctuations by using magnetic-field insensitive states~\cite{ms_clock}.

 As emphasized in the main text, one of the challenges of quantum-information processing is the implementation of quantum logic gates below the fault-tolerance threshold. So far, this has only been achieved for ion optical qubits coupled by a M\o lmer-S\o rensen (MS) scheme~\cite{ms_ft}, which achieved fidelities $\mathcal{F}\approx 0.993$ for two ground-state cooled ions with gate speeds of $t_{\rm f}\approx 0.05$ ms. As a consequence of the vanishingly small spontaneous decay rate on the optical transition, the scheme is completely insensitive to the photon scattering that occurs for hyperfine qubits due to the Raman beams. The remaining sources of noise are caused by off-resonant contributions to the carrier transition, fluctuations of the laser frequency, and changes in the laser intensities. We note that our scheme can overcome  two of these errors reaching similar fidelities $\mathcal{F}\approx0.99-0.999$, while it shares the sensitivity to fluctuations of the spin-phonon laser Rabi frequencies. The advantage of our decoupling protocol becomes more relevant as the ion number is increased, since the sensitivity of the MS scheme to magnetic field noise grows quadratically with the number of ions~\cite{gates_decoherence}. 
 
In the above schemes, most of the  errors are caused by the particular imperfections of the laser beam arrangements. Recently, there has been a growing interest in developing quantum gates that use radio-frequency or microwave radiation~\cite{microwave_gates_wineland,microwave_gates_wunderlich}, which directly overcome the problems due to photon scattering in the Raman configuration, and benefit from the better control over the amplitude and phase of this type of radiation. In~\cite{microwave_gates_wineland}, two-qubit quantum gates for clock states with a fidelity of $\mathcal{F}\approx0.76$ and speeds of $t_{\rm f}\approx0.4$ ms were achieved by means of oscillating magnetic-field gradients  generated in the near-field of microwave currents. 
In this case, the leading source of error corresponds to the contributions of the oscillating gradient to the carrier transition, and effective ac-Zeeman shifts. In~\cite{microwave_gates_wunderlich},  a two-qubit gate with fidelity $\mathcal{F}\approx0.64$ for $t_{\rm f}\approx 8$ ms, has been achieved by means of static magnetic field gradients and radio-frequency radiation. We note that the leading source of noise in this case is caused by the decoherence induced by magnetic field fluctuations. In comparison, our proposal for hyperfine or Zeeman qubits constitutes an hybrid between the laser- and mircrowave-based methods, which can overcome some of these natural errors  and potentially reach the fault-tolerance threshold regime.

\end{document}